\documentclass[twocolumn]{aastex7}

\definecolor{jaredpurple}{RGB}{93, 63, 211}

\usepackage{float}
\usepackage{bookmark} %% generate table of contents sidebar in pdf 
\bookmarksetup{numbered, open,}
\usepackage{enumitem}
\setlist[enumerate]{itemsep=0mm}
\usepackage{hyperref}
\usepackage{subfigure}
\usepackage{multirow}
\usepackage{xspace}
\usepackage{natbib}
\usepackage{xcolor}  
\usepackage{CJK}
\usepackage{soul}
\usepackage{amsfonts}
\usepackage{tikz}
\usepackage{amsmath}
\usepackage{siunitx}  % has the \num command for numbers

\graphicspath{{./}{figures/}}

\newcommand{\eg}{e.g.}
\newcommand{\Nifs}{$^{56}$Ni\xspace}

\newcommand{\kms}{km\,s$^{-1}$\xspace}
\newcommand{\rhounit}{g\,cm$^{-3}$\xspace}
\newcommand{\Msun}{M$_{\odot}$\xspace}
\newcommand{\ergs}{erg\,s$^{-1}$\xspace}
\newcommand{\um}{$\mu m$\xspace}

\newcommand{\alpharho}{$\alpha_{\rho}$\xspace}

\newcommand{\vstart}{$v_\textrm{start}$\xspace}
\newcommand{\vinner}{$v_\textrm{inner}$\xspace}
\newcommand{\vouter}{$v_\textrm{outer}$\xspace}
\newcommand{\vzeroCa}{$v_{X_\textrm{Ca}=0}$\xspace}
\newcommand{\Tinner}{$T_\textrm{inner}$\xspace}
\newcommand{\fsigma}{$f_\sigma$\xspace}

\newcommand{\tardisversionname}{2025.03.23}
\newcommand{\tardis}{\textsc{tardis}\xspace}
\newcommand{\ultranest}{\textsc{ultranest}\xspace}

\newcommand{\Obrienetal}{J.~T.~O'Brien et al.}
\newcommand{\vinnervalue}{9164}

\received{12/25/2025} 
\revised{1/23/2026}
\accepted{3/23/2026}
\submitjournal{ApJL}
\shorttitle{Spectral Inference of SN~2014L}
\shortauthors{Lu et al.}

\begin{document}
\begin{CJK*}{UTF8}{gbsn}

\title{Traces of Helium Detected in Type Ic Supernova 2014L}

\correspondingauthor{Jing Lu}
\email{lujing8@msu.edu}

\author[0000-0002-3900-1452]{Jing Lu (陆晶)}
\affil{Department of Physics and Astronomy, Michigan State University, East Lansing, MI 48824, USA}
\email{lujing8@msu.edu}

\author[0000-0002-0479-7235]{Wolfgang E. Kerzendorf}
\affil{Department of Physics and Astronomy, Michigan State University, East Lansing, MI 48824, USA}
\affil{Department of Computational Mathematics, Science, and Engineering, Michigan State University, East Lansing, MI 48824, USA}
\email{wkerzend@msu.edu}

\author[0000-0003-3615-9593]{John T. O'Brien}
\affil{Department of Astronomy University of Illinois Urbana-Champaign Champaign, Illinois 61801-3633, USA}
\email{jobrien585@gmail.com}

\author[0000-0001-7132-0333]{Maryam Modjaz}
\affil{Department of Astronomy, University of Virginia, Charlottesville, VA 22904, USA}
\email{vru7qe@virginia.edu}

\author[0000-0003-1012-3031]{Jared A. Goldberg}
\affil{Department of Physics and Astronomy, Michigan State University, East Lansing, MI 48824, USA}
\affil{Center for Computational Astrophysics, Flatiron Institute, 162 5th Avenue, New York, NY 10010, USA}
\email{goldstar@msu.edu}

\author[0000-0001-7571-0742]{Nutan Chen}
\affil{Agile Robots AG, Plinganserstrasse 134, Munich 81369, Germany \footnote{During this work, Nutan Chen was affiliated with the Machine \\ Learning Research Lab at Volkswagen
Group}}
\email{nutan.chen@gmail.com}

\author[0009-0001-8470-275X]{Erin Visser}
\affil{Department of Physics and Astronomy, Michigan State University, East Lansing, MI 48824, USA}
\email{visserer@msu.edu}

\author[0000-0002-1560-5286]{Joshua V. Shields}
\affil{Department of Physics and Astronomy, Michigan State University, East Lansing, MI 48824, USA}
\email{shield90@msu.edu}

\author[0000-0001-7343-1678]{Andrew G. Fullard}
\affil{Department of Physics and Astronomy, Michigan State University, East Lansing, MI 48824, USA}
\email{fullarda@msu.edu}

%%%%%%%%%%%%%%%%%%%%%%%%%%%%%%%%%%%%%%%%%%%%%%%%%%%%%%%%%%%%%%%%%%%%%%%%%%%%%%%%
%%%%%%%%%%%%%%%%%%%%%%%%%%%%%%%%%%%%%%%%%%%%%%%%%%%%%%%%%%%%%%%%%%%%%%%%%%%%%%%%
\begin{abstract}
The absence of helium features in optical spectra is one of the classification criteria for Type Ic supernovae (SNe~Ic). 
However, it is highly debated whether helium is truly absent in ejecta or spectroscopically undetectable in the optical region.
The near-infrared (NIR) region contains cleaner He lines that are less blended with other common ions in SNe~Ic ejecta. 
We perform full spectral modeling on the near-peak-light optical and NIR spectra of the SN~Ic 2014L to quantitatively constrain helium and other outer-ejecta properties, using the radiative transfer code \tardis.  
We employ a deep-learning emulator for SNe Ic spectra that serves as a fast surrogate for \tardis simulations. We then integrate the emulator within the Bayesian inference framework to infer the ejecta properties.
The emulator achieves a mean fractional error of 1\% between the emulated and \tardis fluxes across all wavelengths and all samples in the test dataset. 
We constrain 0.018 to 0.020~\Msun (16\% to 84\% posterior percentile) of He above the photosphere near peak light in SN~2014L, inferred from the observed spectra covering 3500~\AA\ to \num{24000}~\AA. 
A Bayesian statistical test shows that the observed spectra are inconsistent with no helium. 
Furthermore, the posterior favors a power-law density exponent of $-7.04$ to $-6.88$ (16\% to 84\% credible interval), consistent with theoretical calculations of radiation-dominated explosions. 
This work demonstrates that Bayesian radiative-transfer inference over a wide wavelength range provides a powerful path toward systematic constraints on He in SNe~Ic.

\end{abstract}

%%%%%%%%%%%%%%%%%%%%%%%%%%%%%%%%%%%%%%%%%%%%%%%%%%%%%%%%%%%%%%%%%%%%%%%%%%%%%%%%
\section{Introduction}  \label{sec:intro}

%%%% Intro of SESN and lead into the He problem 
Stripped-envelope (SE) supernovae (SNe) are core-collapse explosions of massive stars that seemingly have lost part or all of their outer H/He layers before death \citep{filippenko_optical_1997, gal-yam_observational_2017, modjaz_new_2019}. 
The outer envelope can be removed through metallicity-dependent line-driven winds \citep[\eg, ][]{woosley_evolution_1993, heger_how_2003, crowther_physical_2007} and/or binary interaction such as Roche–lobe overflow and common-envelope evolution \citep[\eg, ][]{wheeler_peculiar_1985, podsiadlowski_presupernova_1992, sana_binary_2012, eldridge_death_2013, fang_hybrid_2019, sun_uv_2023}. 
Recent studies have shown growing evidence that intermediate-mass binaries are the dominant progenitor channel for SESNe \citep[see][and references within]{ercolino_interacting_2023, solar_binary_2024, zapartas_demographics_2025}. 
However, stellar simulations of such systems predict a minimum of a few $10^{-1}$~\Msun residual He \citep[see][and references within]{yoon_evolutionary_2015}, which makes it difficult to explain SNe~Ic with binary stripping.

%%%% He might be present in the NIR 
It has long been debated whether the absence of He features in the optical spectra of SNe~Ic truly indicates a lack of He in their ejecta \citep[\eg][]{clocchiatti_sn_1996, taubenberger_sn_2006, hunter_extensive_2009, piro_transparent_2014, liu_analyzing_2016, modjaz_spectral_2016}. 
Spectral modeling works have shown that a trace amount of He could produce clearly-identifiable features in the near-infrared (NIR) but not in optical \citep[\eg][]{teffs_type_2020, lu_physics-driven_2025}.
Early studies on NIR spectra of SNe~Ic proposed that the strong absorption feature near 1~\um could arise from the \ion{He}{1} 1.083~\um transition \citep{wheeler_sn_1994, filippenko_type_1995}. 
Subsequent spectral modeling, however, showed that this feature potentially is not attributed to \ion{He}{1} alone and might instead be produced by a blend of ions, such as \ion{C}{1}, \ion{Si}{1} and \ion{Mg}{2} \citep[\eg][]{millard_direct_1999, baron_spectral_1999, sauer_properties_2006, valenti_carbon-rich_2008, williamson_modeling_2021}. 
More recently, a comprehensive NIR sample study of SESNe found that roughly half of their SN~Ic sample exhibits a weak \ion{He}{1} 2.058~\um line, suggesting that \ion{He}{1} 1.083~\um may contribute to the strong 1~\um feature as well \citep{shahbandeh_carnegie_2022}. 
Precise He mass measurements are crucial for determining whether SNe~Ic are genuinely He-free or whether binary stellar models underpredict the extent of He stripping. 
Indeed, the Helium question also strongly impacts other stripped SN-related explosions, such as Superluminous SNe Ic (e.g. \citealt{kumar_detection_2025}) and Ca-Strong Transients (e.g., \citealt{Kumar26_CaStrong}).

%%%%%% Why modeling He is hard and previous results
A common approach to estimating the He mass involves comparing observational data with synthetic spectra produced by radiative-transfer simulations.
Because typical radiative temperatures in supernova ejecta are insufficient to thermally ionize He, He can remain spectroscopically hidden unless non-thermal ionization/excitation from high-energy particles is efficient in the line-forming region \citep{harkness_early_1987, lucy_nonthermal_1991, kozma_gamma-ray_1992}. 
Only a few radiative-transfer codes can thus model He reliably, as this requires a non-local-thermodynamic-equilibrium (NLTE) treatment \citep[\eg][]{hachinger_how_2012, dessart_nature_2012, boyle_helium_2017}. 

%%%%%%% previous He mass estimation 
Previous studies have reached differing conclusions about how much He can remain undetected, with reported upper limits spanning $10^{-2} - 10^{0}$~\Msun\ for different SESNe \citep{hachinger_how_2012, dessart_nature_2012, dessart_one-dimensional_2015, piro_transparent_2014, williamson_modeling_2021, lu_physics-driven_2025}. 
These works consistently demonstrate that the detectability of He does not depend on He mass alone but also on the density structure, composition, and radiation field of the ejecta.
As a result, quantifying He mass requires full spectral modeling with radiative-transfer simulations that self-consistently treat these coupled physical processes. 

%%%%% Need for Bayesian inference and emulation
However, these previous analyses rely on limited model sets that explore only a narrow region of the relevant parameter space. 
Robust He constraints require a Bayesian framework that quantifies uncertainties and captures parameter degeneracies by jointly analyzing the optical and NIR spectra. 
Such inference requires millions of spectral evaluations, far more than can be computed with direct radiative-transfer simulations within any practical timeframe \citep{kerzendorf_dalek_2021}.
Neural-network emulators address this computational bottleneck by acting as fast surrogates for radiative-transfer models, enabling efficient exploration of the large parameter space while retaining sufficient physical fidelity \citep[\eg][]{vogl_spectral_2020, kerzendorf_dalek_2021, kerzendorf_probabilistic_2022, fullard_new_2022, obrien_1991t-like_2024, vogl_no_2024, karthik_yadavalli_radiative_2025}.
This emulator-accelerated inference approach allows systematic exploration over a broader range of physical conditions and reduces biases associated with incomplete model coverage.

%%%%% This work 
In this work, we utilize both optical and NIR spectra of a SN~Ic object SN~2014L \citep{zhang_optical_2018, shahbandeh_carnegie_2022} as a case study, adapting a Bayesian inference approach developed by \citet{obrien_probabilistic_2021, obrien_1991t-like_2024}.
This inference approach can determine the probability distribution of the ejecta properties. 
In this work, we use the open-source radiative transfer code \tardis \citep{kerzendorf_spectral_2014} with the He treatment developed by \citet{boyle_helium_2017}, which is an analytical approximation of the numerical NLTE calculations from \citet{hachinger_how_2012}. 
%%%%% Paper outline. 
In Section~\ref{sec:obs data}, we summarize the observational data of SN~2014L that form the basis of our inference. 
Section~\ref{sec:methods} describes the outer ejecta model, the \tardis emulator training, and the Bayesian inference framework. 
We present and discuss the results in Section~\ref{sec:results}, and we summarize the main conclusions in Section~\ref{sec:conclusion}.

%%%%%%%%%%%%%%%%%%%%%%%%%%%%%%%%%%%%%%%%%%%%%%%%%%%%%%%%%%%%%%%%%%%%%%%%%%%%%%%%
%%%%%%%%%%%%%%%%%%%%%%%%%%%%%%%%%%%%%%%%%%%%%%%%%%%%%%%%%%%%%%%%%%%%%%%%%%%%%%%%
\section{Observational Data: SN~2014L} \label{sec:obs data}
%%%%% The object we choose and the major stats 
We model the observed spectra of SN~2014L, which is classified as SN~Ic based on its optical properties \citep{zhang_optical_2018}. 
We adopt MJD$= 56693.86$ as the UBVRI quasi-bolometric light-curve peak from \citet{zhang_optical_2018}. 
SN~2014L was first detected on MJD$= 56682.8$ by Koichi Itagaki at the Takamizawa station, implying a lower limit of 13~days for the rise time. 
Unless otherwise noted, we refer to the peak as the maximum of the UBVRI quasi-bolometric light curve in this work.

%%%%% The details of the chosen spectra
While the published dataset on SN~2014L contains six optical spectra \citep{zhang_optical_2018} and two NIR spectra \citep{shahbandeh_carnegie_2022} before 10~days past peak, for our theoretical work, we choose one optical and one NIR spectrum, both obtained near peak light. 
We adopt the earliest available NIR spectrum from \citet{shahbandeh_carnegie_2022}, taken on MJD~56695.9 (2~days after peak) with the FIRE instrument on the 6.5~m Magellan Baade Telescope at Las Campanas Observatory, as part of the Carnegie Supernova Project~II \citep{Phillips2019, Hsiao2019}. 
The optical spectrum is taken from \citet{zhang_optical_2018}, obtained with the YFOSC instrument on the 2.4~m Li-Jiang Telescope at Yunnan Observatories on MJD~56693.9 (0~days relative to peak).

%%%%%%% wavelength masking 
For the optical spectrum, we trim $\sim$50~\AA\ at each edge in the rest frame of SN~2014L to limit instrumental distortions, and we mask out narrow host-galaxy emission lines. 
For the NIR spectrum, we trim $\sim$10~\AA\ on the blue side and an extended $\sim$1800~\AA\ on the red side, where the data are strongly affected by thermal background \citep{simcoe_fire_2013}.
Additionally, we mask out wavelength regions that are strongly affected by the telluric absorptions in the NIR.

%%%%%%%%%%%%%%%%%%%%%%%%%%%%%%%%%%%%%%%%
%%%%%%%%%%%%%%%%%%%%%%%%%%%%%%%%%%%%%%%%
\begin{figure*}[htb!]
\centering
\includegraphics[width=\textwidth]{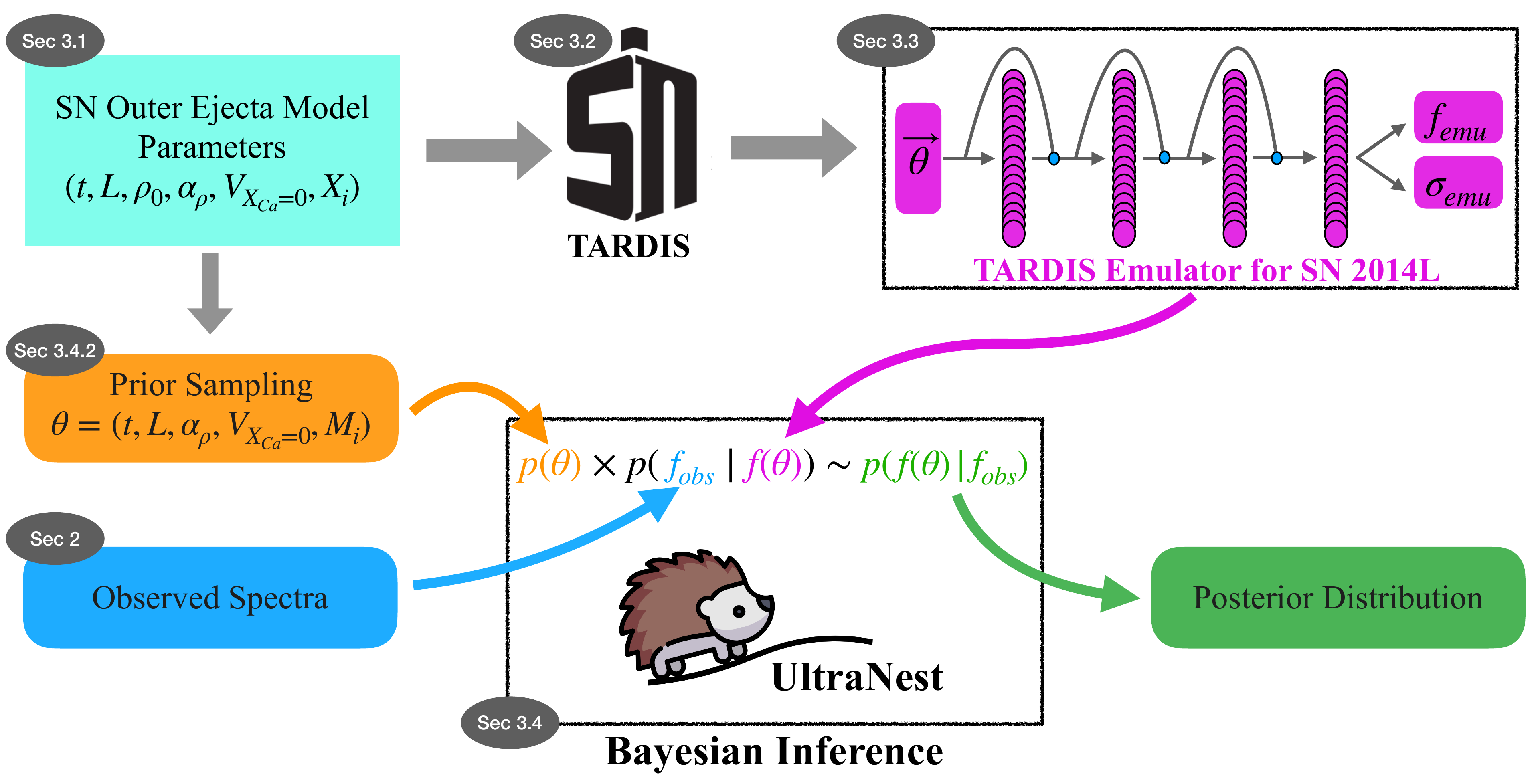}
\caption{Flowchart of the spectral inference methodology applied in this work, adapted from \citet{obrien_probabilistic_2021} and \citet{obrien_1991t-like_2024}. The corresponding section of each component is labeled.}
\label{fig: methods - flowchart}
\end{figure*}
%%%%%%%%%%%%%%%%%%%%%%%%%%%%%%%%%%%%%%%%
%%%%%%%%%%%%%%%%%%%%%%%%%%%%%%%%%%%%%%%%

%%%%%%%%%%%%%%%%%%%%%%%%%%%%%%%%%%%%%%%%%%%%%%%%%%%%%%%%%%%%%%%%%%%%%%%%%%%%%%%%
%%%%%%%%%%%%%%%%%%%%%%%%%%%%%%%%%%%%%%%%%%%%%%%%%%%%%%%%%%%%%%%%%%%%%%%%%%%%%%%%
\section{Methods} \label{sec:methods}
%%%%%% Section intro - general
In this Section, we present the workflow we use to infer the outer ejecta properties of the SN Ic 2014L. 
Our methodology follows the inference frameworks of \citet{obrien_probabilistic_2021} and \citet{obrien_1991t-like_2024}, in which \tardis \citep{kerzendorf_spectral_2014} is used for inference via a probabilistic emulator \citep{kerzendorf_probabilistic_2022}.
A schematic overview of the workflow is presented in Figure~\ref{fig: methods - flowchart}. 
The relevant source code and configuration files are publicly available in Zenodo\footnote{\url{https://doi.org/10.5281/zenodo.19394486}}. 

%%%%%% Section intro - details of each stage
Section~\ref{subsec: method - SN model} introduces the parameterization of the outer ejecta model.
In Section~\ref{subsec: method - tardis}, we describe the detailed configuration of the \tardis radiative transfer code. 
Section~\ref{subsec: method - emulator} details the construction of the radiative transfer emulator. 
Finally, we present the model inference method in Section~\ref{subsec: method - inference}.

%%%%%%%%%%%%%%%%%%%%%%%%%%%%%%%%%%%%%%%%%%%%%%%%%%%%%%%%%%%%%%%%%%%%%%%%%%%%%%%%
\subsection{Parametrization of the ejecta model} \label{subsec: method - SN model}
%%%%% General claim about the OUTER ejecta modeling
We use a parameterized model to describe the ejecta above the photosphere of SN~2014L at an epoch near maximum light. 
A detailed list of the model parameters and their corresponding prior ranges is provided in Table~\ref{table: parameters}. 
Below, we outline both the physical motivations for these choices and the adopted distributions.

%%%%% Powerlaw density profile and parameter range. 
The ejecta in core-collapse explosions reach homologous expansion a few days after shock breakout \citep[\eg][]{dessart_core-collapse_2011, tsang_comparing_2020}, during which the radius grows linearly with velocity and time, $r \propto v t$. 
A power-law density profile provides a good approximation of the outer ejecta structure in SESNe hydrodynamic models \citep[\eg][]{iwamoto_theoretical_1994, mazzali_modelling_2017}. 
Thus, in our model, at a chosen time $t$ after the explosion, the density as a function of the velocity is given by
\begin{equation} \label{eq: density}
\rho(v) = \rho_0 \left(\frac{t_0}{t}\right)^3 \left(\frac{v}{v_0}\right)^{\alpha_{\rho}},
\end{equation}
where $\alpha_{\rho}$ is the power-law index and $t_0 = 5$~day, $v_0 =$\num{5000}~\kms, and $\rho_0$ define normalization at the reference point. 
Motivated by previous SESN studies \citep[\eg][]{farmer_nucleosynthesis_2023, hachinger_how_2012, teffs_type_2020, teffs_observations_2021, williamson_modeling_2021}, we sample $\alpha_{\rho}$ uniformly between $-10$ and $-6$, and $\rho_{0}$ log-uniformly between $3.16\times10^{-11}$~\rhounit\ and $1\times10^{-8}$~\rhounit.

%%%%% Choice of elements and velocity distribution
Following the SN~Ic models of \citet{hachinger_how_2012}, we include 13 elements whose mass fractions exceed $10^{-5}$ in layers above \num{5000}~\kms: He, C, O, Ne, Na, Mg, Si, S, Ca, Ti, Cr, Fe, and $^{56}$Ni. 
We adopt a uniform composition for most of the elements above the photosphere to model the near-peak spectra, motivated by previous studies showing that the outer ejecta composition in a subset of SESN models is nearly uniform \citep[\eg,][]{teffs_type_2020,teffs_observations_2021}.
Introducing a more stratified abundance structure will greatly increase the dimensionality of the parameter space and more complex prior sampling behavior.

During preliminary tests, however, we found that assuming a uniform Ca abundance leads to excessively broad, high-velocity Ca features that are inconsistent with the observed spectra. Hence, we model Ca using a two-zone abundance structure, in which the outer zone is depleted of Ca, analogous to the reduced outer-layer Ca treatment adopted by \citet{hachinger_how_2012} for SESNe modeling.
This transition velocity, \vzeroCa, is sampled from a uniform distribution between \num{10000}~\kms and the outer boundary velocity (\vouter). 
Initial tests showed that a \vouter above \num{35000}~\kms does not affect the spectral features significantly in our parameter space. 
We therefore fix \vouter $=$\num{35000}~\kms for the rest of this work.

%%%%% Mass fractions 
We sample the mass fraction of each element logarithmically in a range informed by previous SESN modeling studies \citep[\eg][]{farmer_nucleosynthesis_2023, hachinger_how_2012, teffs_type_2020, teffs_observations_2021, williamson_modeling_2021}. 
We use O as a ``filler'' element after sampling all other elemental mass fractions so that the sum of mass fractions is one, which is a common practice in SN modeling since their spectrum is relatively insensitive to O abundance \citep[\eg,][]{hachinger_analysis_2011, hachinger_type_2017, obrien_probabilistic_2021}. 
The full ranges of these elemental mass fractions are listed in Table~\ref{table: parameters}.

%%%%% Supernova time and luminosity
\citet{zhang_optical_2018} estimate the peak UVOIR luminosity of SN~2014L to be $L = (2.06 \pm 0.05)\times10^{42}$~erg\,s$^{-1}$ using multiband photometry and a distance of $13.9 \pm 1.5$~Mpc. 
By fitting the early-time light curves with a power-law model, \citet{zhang_optical_2018} obtains a rise time of approximately 13~days from shock breakout to peak, with an uncertainty of 10--20\%. 
Guided by these estimates, but adopting a more conservative range for inference, we sample $L$ log-uniformly between $6.31 \times 10^{41}$ and $3.98 \times 10^{42}$~\ergs, and we sample the time since explosion ($t$) for the optical spectrum at peak uniformly between 8 and 36~days.

%%%%%% Bridge into TARDIS, mention the conversion of the parameters 
Using the parameter distributions described above, we randomly generate approximately \num{700000} parameterized outer ejecta models and compute their corresponding synthetic spectra with \tardis. 
These spectra form the training set for the radiative transfer emulator.  
For emulator training and inference, we convert elemental mass fractions into elemental masses using $\rho_0$ and a fixed velocity range, detailed in Section~\ref{subsec: inference - prior}.

%%%%%%%%%%%%%%%%%%%%%%%%%%%%%%%%%%%%%%%%%%
%%%%%%%%%%%%%%%%%%%%%%%%%%%%%%%%%%%%%%%%%%
%%%%%%%%%%%%%%%%%%%%%%%%%%%%%%%%%%%%%%%%%%
%%%%%%%%%%The Parameter table %%%%%%%%%%%%
\begin{table*}[thb!]
\centering
\caption{Model parameters and prior distributions} \label{table: parameters}
\begin{tabular}{c||lcc||lcc}
\hline \hline
                                       & \multicolumn{3}{c||}{\tardis Sample Grid\tablenotemark{a}}            & \multicolumn{3}{c}{Inference Prior}                                                                                                                      \\ \hline
                                       & \multicolumn{1}{c|}{Parameter}               & \multicolumn{1}{c|}{Distribution} & Range          & \multicolumn{1}{c|}{Parameter}                  & \multicolumn{1}{c|}{Distribution}             & Range                \\ \hline
\multirow{2}{*}{SN} 
                                       & \multicolumn{1}{l|}{$t$ [days]}              & \multicolumn{1}{c|}{Uniform}      & [8, 36]              & \multicolumn{1}{l|}{$t$ [days]}                 & \multicolumn{1}{c|}{Uniform}                  & [8, 36]              \\
                                       & \multicolumn{1}{l|}{$L$ [erg s$^{-1}$]}             & \multicolumn{1}{c|}{Log-uniform}  & [6.31e+41, 3.98e+42] & \multicolumn{1}{l|}{$L$ [erg s$^{-1}$]}                & \multicolumn{1}{c|}{Log-uniform}              & [6.31e+41, 3.98e+42] \\ \hline
% \multirow{3}{*}{Density\tablenotemark{b}}                            
%                                        & \multicolumn{1}{l|}{\vouter [km s$^{-1}$]}          & \multicolumn{1}{c|}{Uniform}      & [2.5e4, 8e4]  & \multicolumn{1}{l|}{\multirow{3}{*}{\alpharho\tablenotemark{c}}} & \multicolumn{1}{c|}{\multirow{3}{*}{Uniform}} & \multirow{3}{*}{[-10, -6]}  \\
\multirow{2}{*}{Density\tablenotemark{b}}                            
                                       & \multicolumn{1}{l|}{$\rho_0$ [\rhounit]}     & \multicolumn{1}{c|}{Log-uniform}  & [3.16e-11, 1e-8]     & \multicolumn{1}{l|}{\multirow{2}{*}{\alpharho\tablenotemark{c}}} & \multicolumn{1}{c|}{\multirow{2}{*}{Uniform}} & \multirow{2}{*}{[-10, -6]}  \\
                                       & \multicolumn{1}{l|}{\alpharho}               & \multicolumn{1}{c|}{Uniform}      & [-10, -6]            & \multicolumn{1}{l|}{}             & \multicolumn{1}{c|}{}                  &                            \\ \hline
\multirow{14}{*}{Abundance\tablenotemark{e}} 
                                       & \multicolumn{1}{l|}{\vzeroCa [km s$^{-1}$]}         & \multicolumn{1}{c|}{Uniform}      & [\num{10000},  \vouter]    & \multicolumn{1}{l|}{\vzeroCa [km s$^{-1}$]}        & \multicolumn{1}{c|}{Uniform}                  &  [\num{10000},  \vouter]    \\
                                       & \multicolumn{1}{l|}{$X_{He}$}                 & \multicolumn{1}{c|}{Log-uniform}  &    [$1.0e-5, 5.0e-2$]    & \multicolumn{1}{l|}{$M_{He}$ [M$_{\odot}$]}           & \multicolumn{1}{c|}{Log-uniform}              &  [$3.4e-5, 1.5e+2$]                    \\
                                       & \multicolumn{1}{l|}{$X_{C}$}                  & \multicolumn{1}{c|}{Log-uniform}  &    [$5.0e-2, 9.0e-1$]    & \multicolumn{1}{l|}{$M_{C}$ [M$_{\odot}$]}            & \multicolumn{1}{c|}{Log-uniform}              &  [$1.6e-1, 2.8e+3$]                   \\
                                       & \multicolumn{1}{l|}{$X_{O}$\tablenotemark{d}} & \multicolumn{1}{c|}{Log-uniform}  &    [$4.6e-7, 9.4e-1$]    & \multicolumn{1}{l|}{$M_{O}$ [M$_{\odot}$]}            & \multicolumn{1}{c|}{Log-uniform}              &  [$2.0e-5, 2.8e+3$]                   \\
                                       & \multicolumn{1}{l|}{$X_{Ne}$}                 & \multicolumn{1}{c|}{Log-uniform}  &    [$1.0e-2, 2.5e-1$]    & \multicolumn{1}{l|}{$M_{Ne}$ [M$_{\odot}$]}           & \multicolumn{1}{c|}{Log-uniform}              &  [$3.2e-2, 7.6e+2$]                   \\
                                       & \multicolumn{1}{l|}{$X_{Na}$}                 & \multicolumn{1}{c|}{Log-uniform}  &    [$1.0e-4, 2.0e-2$]    & \multicolumn{1}{l|}{$M_{Na}$ [M$_{\odot}$]}           & \multicolumn{1}{c|}{Log-uniform}              &  [$3.2e-4, 6.0e+1$]                   \\
                                       & \multicolumn{1}{l|}{$X_{Mg}$}                 & \multicolumn{1}{c|}{Log-uniform}  &    [$5.0e-5, 1.0e-2$]    & \multicolumn{1}{l|}{$M_{Mg}$ [M$_{\odot}$]}           & \multicolumn{1}{c|}{Log-uniform}              &  [$1.7e-4, 3.1e+1$]                   \\
                                       & \multicolumn{1}{l|}{$X_{Si}$}                 & \multicolumn{1}{c|}{Log-uniform}  &    [$5.0e-5, 1.0e-2$]    & \multicolumn{1}{l|}{$M_{Si}$ [M$_{\odot}$]}           & \multicolumn{1}{c|}{Log-uniform}              &  [$1.6e-4, 6.0e+1$]                   \\
                                       & \multicolumn{1}{l|}{$X_{S}$}                  & \multicolumn{1}{c|}{Log-uniform}  &    [$5.0e-5, 1.0e-2$]    & \multicolumn{1}{l|}{$M_{S}$ [M$_{\odot}$]}            & \multicolumn{1}{c|}{Log-uniform}              &  [$1.6e-4, 6.3e+1$]                   \\
                                       & \multicolumn{1}{l|}{$X_{Ca}$}                 & \multicolumn{1}{c|}{Log-uniform}  &    [$5.0e-5, 1.0e-2$]    & \multicolumn{1}{l|}{$M_{Ca}$ [M$_{\odot}$]}           & \multicolumn{1}{c|}{Log-uniform}              &  [$1.6e-4, 6.2e+1$]                   \\
                                       & \multicolumn{1}{l|}{$X_{Ti}$}                 & \multicolumn{1}{c|}{Log-uniform}  &    [$1.0e-6, 3.0e-3$]    & \multicolumn{1}{l|}{$M_{Ti}$ [M$_{\odot}$]}           & \multicolumn{1}{c|}{Log-uniform}              &  [$3.3e-6, 8.8e+0$]                   \\
                                       & \multicolumn{1}{l|}{$X_{Cr}$}                 & \multicolumn{1}{c|}{Log-uniform}  &    [$1.0e-6, 3.0e-3$]    & \multicolumn{1}{l|}{$M_{Cr}$ [M$_{\odot}$]}           & \multicolumn{1}{c|}{Log-uniform}              &  [$3.5e-6, 8.9e+0$]                   \\
                                       & \multicolumn{1}{l|}{$X_{Fe}$}                 & \multicolumn{1}{c|}{Log-uniform}  &    [$1.0e-5, 1.0e-2$]    & \multicolumn{1}{l|}{$M_{Fe}$ [M$_{\odot}$]}           & \multicolumn{1}{c|}{Log-uniform}              &  [$3.2e-5, 2.9e+1$]                   \\
                                       & \multicolumn{1}{l|}{$X_{^{56}Ni}$}            & \multicolumn{1}{c|}{Log-uniform}  &    [$1.0e-5, 1.0e-2$]    & \multicolumn{1}{l|}{$M_{^{56}Ni}$ [M$_{\odot}$]}      & \multicolumn{1}{c|}{Log-uniform}              &  [$3.4e-5, 3.0e+1$]                   \\ \hline
Others                                 & \multicolumn{3}{l||}{}                                                                                       & \multicolumn{1}{l|}{\fsigma}                          & \multicolumn{1}{c|}{Log-uniform}              &  [$1.0e-6, 1.0e-1$]                   \\
\hline \hline     
\end{tabular}
\tablenotetext{a}{The inner boundary velocity \vinner is solved iteratively using the \texttt{v\_inner\_solver} workflow in \tardis developed in \Obrienetal\ (in prep.).}
\tablenotetext{b}{The reference time and velocity for the density profile are fixed at $t_0 = 5$~day and $v_0 =$ \num{5000}~\kms in this work.}
\tablenotetext{c}{We fix \vouter at \num{35000}~\kms, and $\rho_0$ is determined from the total elemental mass based on Eq.~\ref{eq: total mass}.}
\tablenotetext{d}{Mass fraction of O is used as a filler element after sampling all other elemental mass fractions.}
\tablenotetext{e}{The elemental masses used for inference represent the integrated mass from \vstart to \vouter following Eq.~\ref{eq: total mass}} given a density profile, which is not the ejecta mass.
\end{table*}
%%%%%%%%%%%%%%%%%%%%%%%%%%%%%%%%%%%%%%%%%%
%%%%%%%%%%%%%%%%%%%%%%%%%%%%%%%%%%%%%%%%%%

%%%%%%%%%%%%%%%%%%%%%%%%%%%%%%%%%%%%%%%%%%%%%%%%%%%%%%%%%%%%%%%%%%%%%%%%%%%%%%%%
\subsection{Radiative Transfer Code: \tardis} \label{subsec: method - tardis}
%%%%%% Introduce TARDIS 
We compute synthetic spectra for the outer ejecta models using the radiative transfer code \tardis\footnote{\url{https://github.com/tardis-sn/tardis}} \citep{kerzendorf_spectral_2014}. 
The version of \tardis employed here, version \tardisversionname\footnote{\url{https://github.com/tardis-sn/tardis/tree/release-2025.03.23}} \citep{kerzendorf_tardis-sntardis_2025},  solves for the steady-state radiation field as well as the plasma conditions for a homologously expanding ejecta at a specified luminosity and time since explosion. 
\tardis adopts a photospheric inner boundary and iteratively determines the radiation field by propagating indivisible energy packets through a spherically symmetric ejecta. 
In general, we describe the plasma state with an NLTE approximation called dilute radiation field \citep{kerzendorf_spectral_2014}. 
Additionally, for the non-thermal excitation of He by fast electrons, we use the methodology described in \citet{boyle_helium_2017}. 
We use configuration choices consistent with previous \tardis studies of SESNe \citep[see e.g.][]{williamson_modeling_2021, lu_physics-driven_2025}. 
The key settings adopted in this work are summarized in Table~\ref{table: tardis_settings}, and for further details on the assumptions and implementations in tardis, see \citet{kerzendorf_spectral_2014} and \citet{boyle_helium_2017}.

%%%%%%%%%%%%%%%%%%%%%%%%%%%%%%%%%%%
%%%%%%%%%%%%%%%%%%%%%%%%%%%%%%%%%%%
\begin{deluxetable}{l|l}[htb!]
\centering
\tablecaption{Key configuration in \tardis simulations \label{table: tardis_settings}}
\tablehead{
\colhead{Configuration name} & \colhead{Setting}}
\startdata
Atomic data          & kurucz\_cd23\_chianti\_H\_He.h5 \\
Ionization           & nebular                     \\
Excitation           & dilute-lte                  \\
Line interaction     & macroatom                   \\
Helium treatment     & recomb-nlte                 \\
\enddata
\end{deluxetable}
%%%%%%%%%%%%%%%%%%%%%%%%%%%%%%%%%%%
%%%%%%%%%%%%%%%%%%%%%%%%%%%%%%%%%%%

%%%%%% The dynamic velocity grid - mainly the v_inner solver workflow
We employ the \texttt{v\_inner\_solver} workflow developed by \Obrienetal\ (in prep.) to iteratively determine the inner boundary velocity (\vinner). 
This workflow iteratively identifies the velocity at which the Rosseland mean optical depth reaches the target value alongside the standard \tardis radiation-field and plasma determinations. 
In this work, we set the target Rosseland mean optical depth to $2/3$. 
To ensure that the solver can locate the appropriate \vinner, we initialize the velocity grid at a low value (\vstart $= \num{3000}$~\kms). 
We construct the velocity grid using logarithmically spaced shells, which provide higher resolution in dense inner regions of the ejecta.

%%%%%%%%%%%%%%%%%%%%%%%%%%%%%%%%%%%%%%%%%%%%%%%%%%%%%%%%%%%%%%%%%%%%%%%%%%%%%%%%
\subsection{Radiative Transfer Emulator} \label{subsec: method - emulator}
%%%%%% Introduce the prob-Dalek 
We adopt the \texttt{Probabilistic Dalek} architecture \citep{kerzendorf_probabilistic_2022}, a deep-learning neural network with local residual connections, to construct the \tardis emulator within the selected parameter space. 
We briefly summarize the architectural design and training strategy in Appendix~\ref{appendix: emulator achitechture}.
All neural-network models are implemented using \texttt{PyTorch} \citep{paszke_pytorch_2019} and \texttt{PyTorch Lightning} \citep{falcon_pytorch_2024}.

%%%%%% Output spectra space 
Before training, we resample the \tardis spectra from a linear wavelength grid onto a logarithmically spaced grid, following \citet{kerzendorf_dalek_2021}. 
A logarithmic wavelength grid enforces a constant $\Delta log(\lambda)$, corresponding to uniform resolution in expansion velocity that scales with $\Delta \lambda / \lambda$.
The resulting grid spans \num{3000} -- \num{24000}~\AA\ with 800 wavelength points, fully covering the masked observational range. 
The converged values of \vinner and \Tinner from each \tardis run are also included as emulator outputs, along with the interpolated luminosity density. 
All inputs and outputs are standardized by subtracting the mean and scaling to unit variance prior to network training \citep{kerzendorf_dalek_2021}.

%%%%%% Normalization and the train/validation/test split 
For luminosity, elemental masses, and luminosity densities, we take the logarithm of each quantity before standardization so that the inputs passed to the emulator training space have distributions that more closely approximate Gaussian or uniform forms.
The dataset is partitioned into training, validation, and testing subsets using an 85\% / 10\% / 5\% split.
The test data, which contains \num{36334} samples, yields a mean fractional error of 1\% in flux comparison with the TARDIS output across all wavelengths and all samples.
We report more details on the emulator performance in Appendix~\ref{appendix: emulator performance}.

%%%%%%%%%%%%%%%%%%%%%%%%%%%%%%%%%%%%%%%%%%%%%%%%%%%%%%%%%%%%%%%%%%%%%%%%%%%%%%%%
\subsection{Model Inference} \label{subsec: method - inference}
%%%%%%% Inference Introduction
We use a Bayesian approach to infer the posterior distribution of the SN ejecta model parameters by comparing the observed spectrum to the synthetic spectra evaluated with the trained emulator. 
We fit the optical and NIR spectra simultaneously by multiplying their likelihoods using a shared base parameter set and applying observationally motivated offsets in time and luminosity for the NIR spectrum given the difference of 2 days between when the optical and the NIR spectra were taken, see Section~\ref{subsec: inference - likelihood}.

%%%%%% Data processing on the observed spectrum
Prior to evaluating the likelihood, we first interpolate the synthetic spectrum onto the masked wavelength sampling grid of the observed spectrum. 
We then normalize the interpolated synthetic flux to match the apparent continuum of the observed spectrum \citep[for details see][]{obrien_1991t-like_2024}, which removes systematic effects that come from an incomplete modelling of the continuum spectral shape (e.g. reddening).

%%%%%%%%%%%%%%%%%%%%%%%%%%%%%%%%%%%%%%%%%%%%%%%%%%%%%%%%%
\subsubsection{Likelihood} \label{subsec: inference - likelihood}

%%%%%%% Likelihood function
Given a model parameter set $\vec{\theta}$, the likelihood ($\mathcal{L}$) of the synthetic spectrum representing the $i^{th}$ observation is calculated through:
{\small
\begin{equation} \label{eq: likelihood individual}
\log \mathcal{L}_i(\vec{\theta})=-\frac{1}{2} \sum_\lambda\left[\left(\frac{\hat{f}_\lambda^{\mathrm{emu}}(\vec{\theta})-f_\lambda^{\mathrm{obs}}}{\sigma_\lambda(\vec{\theta})}\right)^2+\log \left(2 \pi \sigma_\lambda^2(\vec{\theta})\right)\right],
\end{equation}
}
where the uncertainty term is constituted of both the observational and emulator uncertainty: 
\begin{equation} \label{eq: likelihood sigma}
\sigma_\lambda^2(\vec{\theta})=\sigma_{\mathrm{obs}, \lambda}^2+f_\sigma^2 \left(\hat{f}_\lambda^{\mathrm{emu}}(\vec{\theta})\right)^2+\sigma_{\mathrm{emu}. \lambda}^2(\vec{\theta}),
\end{equation}
Here the $f_\lambda^{\mathrm{obs}}$ and $\sigma_{\mathrm{obs,\lambda}}$ are the observed flux and corresponding uncertainty; 
$\hat{f}_\lambda^{\mathrm{emu}}(\vec{\theta})$ and $\sigma_{\mathrm{emu}. \lambda}^2(\vec{\theta})$ is the continuum matched emulator flux and scaled emulator uncertainty evaluated under model parameter $\vec{\theta}$;
and $f_\sigma$ represents a fractional term of the processed emulator flux, which minimizes bias in the posterior distribution due to systematic uncertainties \citep{obrien_1991t-like_2024}.
Following \citet{obrien_1991t-like_2024}, we sample $f_\sigma$ log-uniformly between $10^{-6}$ and $10^{-1}$.  

%%%%%% Coupling between the optical and NIR
Hence, the total likelihood is:
\begin{equation} \label{eq: likelihood total}
\log \mathcal{L}(\vec{\theta}) = \log \mathcal{L}_{\mathrm{optical}}(\vec{\theta}) + \log \mathcal{L}_{\mathrm{NIR}}(\vec{\theta'}),
\end{equation}
where $\vec{\theta'}$ is a modified set of $\vec{\theta}$ that differs only in time and luminosity.
Based on \citet{zhang_optical_2018}, we use a time difference of 2.0~day between the optical and NIR spectra, and a luminosity ratio of $L(t=2d)/L(t=0d) = 0.98$, respectively.

%%%%%%%%%%%%%%%%%%%%%%%%%%%%%%%%%%%%%%%%%%%%%%%%%%%%%%%%%
\subsubsection{Priors} \label{subsec: inference - prior}

%%%%%% Input model parameter transformation - compute elemental masses
We adopt the parameter ranges used for emulator training as the priors for the inference stage. 
Rather than sampling directly in mass-fraction space ($X_i$), where the abundances must satisfy the constraint $\sum_i X_i = 1$, we transform the abundances into elemental masses ($M_i$) and omit $\rho_0$.  
This transformation preserves the number of degrees of freedom while reducing parameter correlations and simplifying the sampling process.
Assuming spherical symmetry, a power-law density profile (Eq.~\ref{eq: density}), and homologous expansion ($r = vt$), the initial density $\rho_0$ is proportional to the total integrated mass:
\begin{equation} \label{eq: total mass}
M_{total} \propto \int \rho_0\left(\frac{v}{v_0}\right)^{\alpha_{\rho}}\left(\frac{t_0}{t}\right)^3 v^2 \cdot t^3 d v,  
\end{equation}
which illustrates that specifying the total mass fully determines $\rho_0$ for a given density profile.

%%%%%%% Refer to the previous section and point to the prior table
In total, we sample 18 parameters during inference: the luminosity $L$, time since explosion $t$, the Ca abundance transition velocity \vzeroCa, the density exponent \alpharho, the 13 elemental masses $M_i$, and the fractional emulator-uncertainty parameter $f_\sigma$ described in Section~\ref{subsec: inference - likelihood}. 
The prior ranges and distributions for all parameters are listed in Table~\ref{table: parameters}.

%%%%%%%%%%%%%%%%%%%%%%%%%%%%%%%%%%%%%%%%%%%%%%%%%%%%%%%%%
\subsubsection{Nested Sampling: \ultranest} \label{subsec: inference - ultranest}

%%%%%%% Introduce Ultranest and the method
We perform Bayesian inference using the \ultranest \citep{buchner_ultranest_2021}, which implements the nested sampling algorithm \citep{skilling_nested_2004, ashton_nested_2022} to compute the Bayesian evidence and generate posterior samples. 
We configure \ultranest with a minimum of 400 live points and use the \texttt{SliceSampler} as the step sampler. 
For convergence, we require that the contribution of the current live points to the remaining unexplored prior volume (the remainder fraction) be less than 1\%.
For our dataset, this criterion is met after approximately \num{14000000} likelihood evaluations.
Using two NVIDIA RTX A4000 GPUs for emulator synthetic spectra evaluation, the inference took 24 hours to complete. In contrast, performing the same inference with direct \textsc{tardis} simulation, which requires $\sim$15 minutes per spectrum over the same parameter space, would be computationally impractical.

%%%%%%%%%%%%%%%%%%%%%%%%%%%%%%%%%%%%%%%%
%%%%%%%%%%%%%%%%%%%%%%%%%%%%%%%%%%%%%%%%
\begin{figure*}[ht!]
\centering
\includegraphics[width=\textwidth]{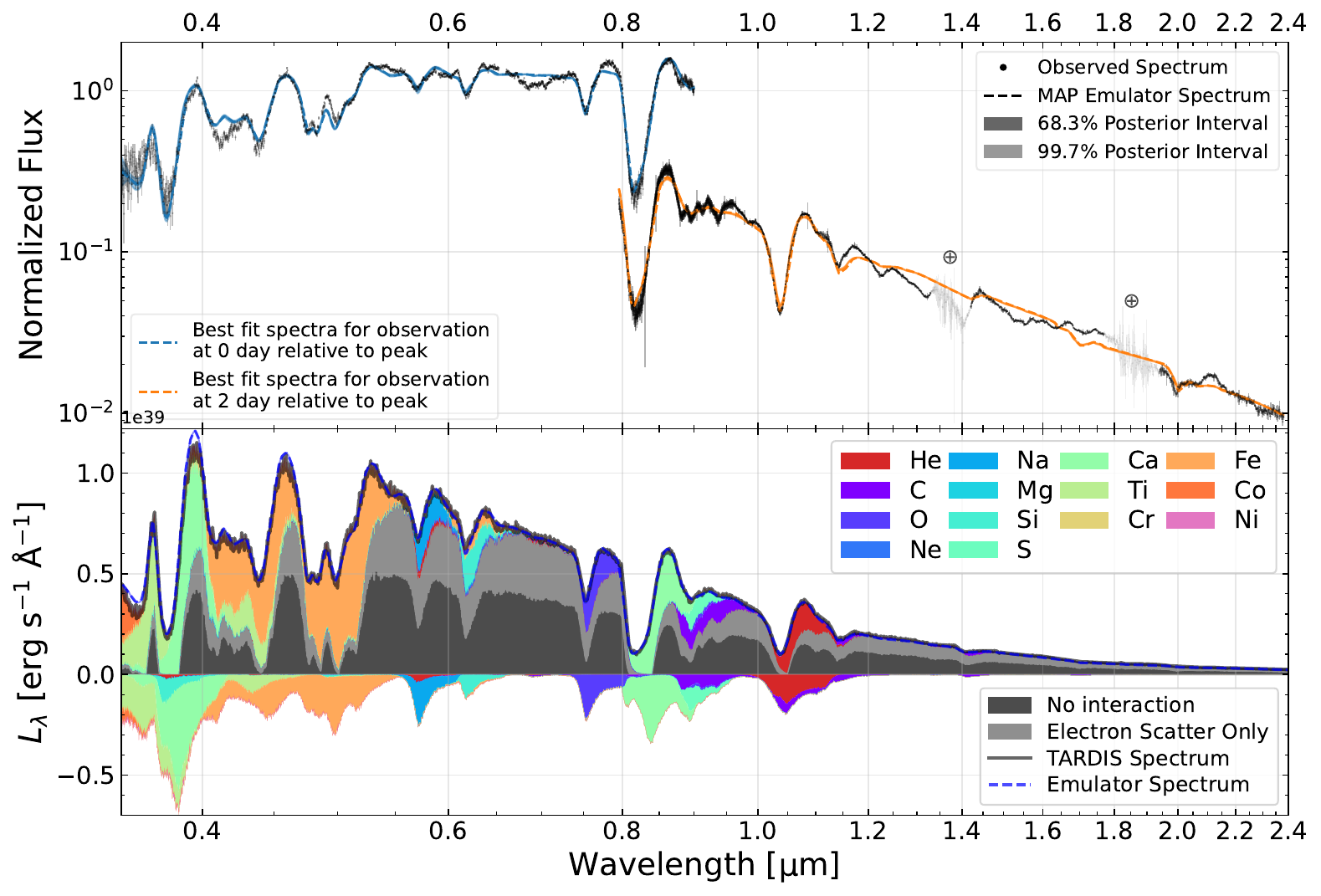}
\caption{
\textit{Top}: The spectral comparison of the observed spectra and the maximum-a-posteriori (MAP) spectra. The observed spectra are plotted as black points with $1\sigma$ error bars. The normalized and continuum-matched emulator spectrum evaluated with the MAP parameter set is plotted in blue and orange for the optical and NIR spectrum, respectively. Note that the two inferred spectra share the same parameters except the luminosity and time, which are locked in a ratio based on observation. The dark and light shaded regions around the MAP indicate the 68.3\% and 99.7\% credible posterior probability intervals, respectively. 
We mark the masked-out telluric regions in the NIR with the $\oplus$ symbol.
\textit{Bottom}: We demonstrate that the emulator spectra (blue dashed line) closely resemble the \tardis simulation (grey solid line) evaluated with the MAP parameter set. The elemental decomposition plot of the \tardis simulation is overlaid, representing the interaction type or ion contribution of the energy packets during the last interaction in the simulation. 
}
\label{fig: posterior_spectral_comparison}
\end{figure*}
%%%%%%%%%%%%%%%%%%%%%%%%%%%%%%%%%%%%%%%%
%%%%%%%%%%%%%%%%%%%%%%%%%%%%%%%%%%%%%%%%

%%%%%%%%%%%%%%%%%%%%%%%%%%%%%%%%%%%%%%%%
%%%%%%%%%%%%%%%%%%%%%%%%%%%%%%%%%%%%%%%%
\begin{figure*}[htb!]
\centering
\includegraphics[width=\textwidth]{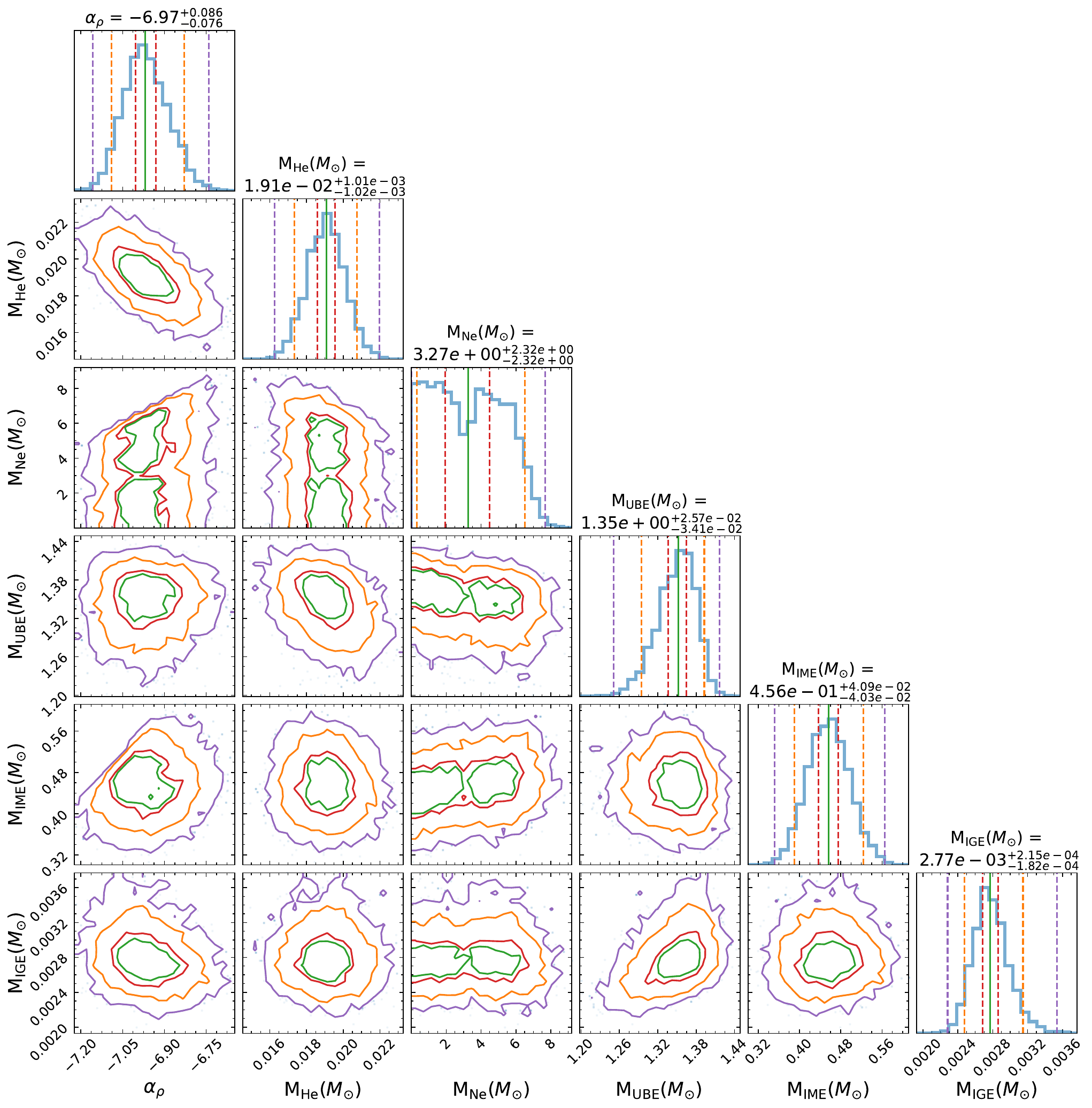}
\caption{The corner plot of the marginalized posterior parameters over \fsigma. The elemental masses presented in this plot are the elemental mass above \vinner in the ejecta.
We group up C and O as the unburnt elements (UBE); Na, Mg, Si, S, and Ca as the intermediate-mass elements (IME); and  
Ti, Cr, Fe, and $^{56}$Ni as the iron-group elements (IGE).
The contour lines denote the 50, 68, 95, and 97$\%$ quantiles and are color-matched to the vertical quantile markers in the diagonal histograms.}
\label{fig: corner_plot_combined_IME_and_IGE}
\end{figure*}
%%%%%%%%%%%%%%%%%%%%%%%%%%%%%%%%%%%%%%%%
%%%%%%%%%%%%%%%%%%%%%%%%%%%%%%%%%%%%%%%%

%%%%%%%%%%%%%%%%%%%%%%%%%%%%%%%%%%%%%%%%%%%%%%%%%%%%%%%%%%%%%%%%%%%%%%%%%%%%%%%%
%%%%%%%%%%%%%%%%%%%%%%%%%%%%%%%%%%%%%%%%%%%%%%%%%%%%%%%%%%%%%%%%%%%%%%%%%%%%%%%%
\section{Results and Discussion} \label{sec:results}
%%%%%% Spectra results overall statement
As shown in Figure~\ref{fig: posterior_spectral_comparison}, our \tardis emulator and inference workflow successfully reproduces the main spectral features near maximum light, including the \ion{Ca}{2}, \ion{Na}{1}, \ion{Si}{2}, and \ion{O}{1} lines, as well as the prominent absorption feature near 1~\um. 
In the bottom panel of Figure~\ref{fig: posterior_spectral_comparison}, we compare the emulator spectrum using the maximum-a-posteriori (MAP) parameter set with the corresponding \tardis simulation,  and display the elemental decomposition of Monte Carlo energy packets. 
The MAP parameter set reproduces the iron-group features between $\lambda$\num{3000}~\AA\ and $\lambda$\num{5500}~\AA\ with only minor discrepancies in line strength. 
In the NIR region, however, the inferred spectra match the observations less accurately than in the optical. 
This limitation likely arises from the assumption of uniform abundances and the use of a wavelength-independent photospheric boundary, as longer wavelengths tend to probe deeper layers of the ejecta \citep[\eg][]{wheeler_explosion_1998, hoflich_infrared_2002}.

%%%%%% Elemental mass grouping
In Figure~\ref{fig: corner_plot_combined_IME_and_IGE}, we plot the pairwise joint posterior distributions of the inferred elemental masses above \vinner. 
For clarity, we group all elements except He and Ne into three categories in the discussion of the results (the inference treats them separately):  
(i) unburnt elements (UBE): C and O;  
(ii) intermediate-mass elements (IME): Na, Mg, Si, S, and Ca; and  
(iii) iron-group elements (IGE): Ti, Cr, Fe, and $^{56}$Ni.  
We treat He separately because it is the primary focus of this study, and we list Ne independently because its posterior distribution remains unconstrained due to the lack of notable Ne features.  
Table~\ref{table: elemental_mass_above_v_inner} reports the individual elemental masses above \vinner, along with their ratios relative to C and Fe. 
We emphasize that these values represent only the portion of each element located above the photosphere and therefore do not correspond to the total ejecta mass.

%%%%%% Section guide
We discuss the inferred luminosity and time since explosion in Section~\ref{subsec: results - Luminoisty and time}, the photospheric properties in Section~\ref{subsec: results - photospheric properties}, the density structure in Section~\ref{subsec: results - density}, and the elemental composition in Section~\ref{subsec: results - composition}.

%%%%%%%%%%%%%%%%%%%%%%%%%%%%%%%%%%%%%%%
\begin{deluxetable*}{c|ccc|ccc|ccc}[htb!]
\centering
\tablecaption{Posterior Percentiles of Elemental Mass above \vinner$=\vinnervalue$~\kms} \label{table: elemental_mass_above_v_inner}
\tablehead{
\multirow{2}{*}{Element} & \multicolumn{3}{c|}{M(\Msun)} & \multicolumn{3}{c|}{X/C} & \multicolumn{3}{c}{X/Fe} \\
& 16\% & 50\% & 84\% 
& 16\% & 50\% & 84\% 
& 16\% & 50\% & 84\% 
}
\startdata
He   & 1.81e-02 & 1.91e-02 & 2.01e-02 & 1.43e-02 & 1.52e-02 & 1.63e-02 & 7.04e+00 & 7.57e+00 & 8.16e+00 \\
C    & 1.22e+00 & 1.25e+00 & 1.27e+00 & $\dots$  & $\dots$  & $\dots$  & 4.77e+02 & 4.97e+02 & 5.18e+02 \\
O    & 9.18e-02 & 9.94e-02 & 1.08e-01 & 7.39e-02 & 7.96e-02 & 8.56e-02 & 3.66e+01 & 3.95e+01 & 4.28e+01 \\
Ne   & 9.53e-01 & 3.28e+00 & 5.59e+00 & 7.56e-01 & 2.63e+00 & 4.51e+00 & 3.81e+02 & 1.29e+03 & 2.22e+03 \\
Na   & 1.06e-02 & 1.22e-02 & 1.37e-02 & 8.46e-03 & 9.75e-03 & 1.10e-02 & 4.20e+00 & 4.83e+00 & 5.50e+00 \\
Mg   & 5.81e-06 & 2.22e-05 & 1.78e-04 & 4.61e-06 & 1.77e-05 & 1.43e-04 & 2.30e-03 & 8.81e-03 & 7.08e-02 \\
Si   & 1.72e-02 & 1.97e-02 & 2.25e-02 & 1.39e-02 & 1.58e-02 & 1.79e-02 & 6.89e+00 & 7.86e+00 & 8.95e+00 \\
S    & 3.84e-01 & 4.24e-01 & 4.64e-01 & 3.07e-01 & 3.39e-01 & 3.73e-01 & 1.53e+02 & 1.68e+02 & 1.85e+02 \\
Ca   & 2.21e-04 & 2.45e-04 & 2.73e-04 & 1.76e-04 & 1.96e-04 & 2.21e-04 & 8.66e-02 & 9.73e-02 & 1.10e-01 \\
Ti   & 1.47e-06 & 1.66e-06 & 1.86e-06 & 1.17e-06 & 1.33e-06 & 1.49e-06 & 5.76e-04 & 6.58e-04 & 7.48e-04 \\
Cr   & 5.56e-08 & 8.70e-08 & 1.54e-07 & 4.46e-08 & 6.96e-08 & 1.23e-07 & 2.22e-05 & 3.47e-05 & 6.15e-05 \\
Fe   & 2.39e-03 & 2.52e-03 & 2.64e-03 & 1.93e-03 & 2.01e-03 & 2.10e-03 & $\dots$  & $\dots$  & $\dots$ \\
Ni56 & 1.43e-04 & 2.35e-04 & 3.93e-04 & 1.16e-04 & 1.89e-04 & 3.14e-04 & 5.79e-02 & 9.37e-02 & 1.54e-01 \\
\enddata
\end{deluxetable*}
%%%%%%%%%%%%%%%%%%%%%%%%%%%%%%%%%%%%%%%

%%%%%%%%%%%%%%%%%%%%%%%%%%%%%%%%%%%%%%%
\subsection{Luminosity and Time} \label{subsec: results - Luminoisty and time}
%%%%%% Disclaimer
We adopt conservative priors for the luminosity and the time since explosion based on the estimates of \citet{zhang_optical_2018} (see Section~\ref{subsec: method - SN model}). 
However, the resulting posterior distributions lie at the edges of these priors for both parameters. 
Below, we discuss plausible interpretations of this behavior.

%%%%%%  Luminosity 
The posterior luminosity of SN~2014L at peak is $\log(L/\mathrm{erg~s^{-1}}) = 42.597^{42.599}_{42.594}$\footnote{We report the 16, 50, and 84\% marginalized posterior percentiles of each parameter in the format of $50\%^{84\%}_{16\%}$.}, approximately a factor of two higher than the UVOIR pseudo-bolometric luminosity reported by \citet{zhang_optical_2018}. 
As noted by \citet{zhang_optical_2018}, the host environment of SN~2014L is likely dusty, with an estimated host color excess ranging from 0.38 to 0.76~mag based on Na~I~D absorption and the $V-R$ color. 
If our inferred luminosity reflects the true UVOIR output, it suggests that SN~2014L may have experienced stronger host or interstellar extinction than previously estimated.

%%%%%%  Time since explosion  
The posterior yields a time since explosion of $35.30^{35.90}_{35.67}$~days for the optical spectrum at peak light. 
\citet{zhang_optical_2018} estimated a rise time of 13~days or 14.5~days from early-time light-curve fits using $t^{1.4}$ and $t^{2}$ power laws, respectively. 
However, the effective rise time can be longer if the ejecta experience an initial expansion phase powered by heating mechanisms in addition to \Nifs\ decay, such as shock heating or energy injection from a central engine, which delays the emergence of the light-curve peak \citep[see][and references therein]{zhang_optical_2018}. 
We note that our parameterized ejecta model assumes homologous expansion beginning at $t=0$. 
Thus, the inferred time since explosion may not correspond to the true rise time if the outer envelope undergoes a non-homologous expansion phase of non-negligible duration.

%%%%%%%%%%%%%%%%%%%%%%%%%%%%%%%%%%%%%%%
\subsection{Photospheric Properties} \label{subsec: results - photospheric properties}
%%%%% Photospheric Properties
From the posterior distribution, we extract the photospheric properties, including \Tinner and \vinner, evaluated using the \tardis emulator trained in Section~\ref{subsec: method - emulator}. 
For the optical spectrum at peak light, the posterior parameter set yields \Tinner $= 6238 \pm 25$~K and \vinner $9164 \pm 43$~\kms.
For the NIR spectrum taken 2~days after peak, the corresponding values are \Tinner $= 6130 \pm 24$~K and \vinner $8992 \pm 43$~\kms. 
These inferred photospheric velocities are consistent with the ion velocities measured by \citet{zhang_optical_2018}, which range from 7650 to \num{14000}~\kms near peak light. 
Our near-peak photospheric properties also agree with the radiative-transfer predictions of He-star progenitor models in \citet{dessart_supernovae_2020}, which exhibit photospheric temperatures of $6000 - 7556$~K and velocities of $8108$ $-$ \num{12876}~\kms (see their Table~3).

%%%%%%%%%%%%%%%%%%%%%%%%%%%%%%%%%%%%%%%
\subsection{Density Profile} \label{subsec: results - density}
%%%%% Power law density exponent 
We infer a density power-law exponent of $\alpha_{\rho} = -6.97^{-6.88}_{-7.04}$ for SN~2014L near peak light. 
Given the uniform prior range of $-10$ to $-6$, this posterior value is consistent with the canonical $\alpha_{\rho} \approx -7$ expected for the outer envelopes of radiation-dominated explosions, as established in theoretical calculations \citep[\eg][]{colgate_early_1969, chevalier_hydrodynamics_1976, chevalier_exploding_1981, iwamoto_theoretical_1994}.

%%%%% Reference density value
In our ejecta model, the initial density value directly reflects the integrated mass of the outer ejecta, see Eq.~\ref{eq: total mass}. 
Because the inferred Ne mass remains unconstrained, the total mass, and therefore the density normalization, also remains unconstrained. 
For this reason, we do not report a value for the initial density. 
This choice does not affect our interpretation of the density exponent, as SESN modeling commonly rescales the density profile while preserving its shape, since these profiles arise from self-similar solutions in explosion models and their absolute normalization scales with the CO-core mass \citep[\eg][]{mazzali_modelling_2017, teffs_how_2020}.

%%%%%%%%%%%%%%%%%%%%%%%%%%%%%%%%%%%%%%%
\subsection{Composition} \label{subsec: results - composition}
%%%%% What the inferred composition represents
We present the posterior composition above the photosphere at peak light in Table~\ref{table: elemental_mass_above_v_inner}. 
%%%%%% Note on uniform abundance 
The steep density gradient and composition present in essentially all realistic SN~Ic progenitors confines the line-forming region to a narrow zone near the photosphere for most elements (with the notable exception of Ca). Thus, variations in the outer layers are not probed in the spectra at this phase. As an example, only 5\% of packets interact at layers above \num{20000}$~\mathrm{km\,s^{-1}}$ in our MAP model. The limited radial sensitivity makes our one-zone abundance description an appropriate representation for this work.

%%%%%% Ne as an unconstrained parameter
In our \tardis simulations, fewer than 1\% of Monte Carlo energy packets interact with Ne. 
Consequently, the inferred Ne mass remains unconstrained, as seen in Figure~\ref{fig: corner_plot_combined_IME_and_IGE}. 
This behavior is consistent with previous SESN modeling studies, in which Ne often acts as a filler element or is assigned a fixed fiducial abundance \citep[\eg][]{mazzali_modelling_2017, ashall_extracting_2020}.
We note that although O serves as a filler element during the construction of the training-sample grid, its posterior distribution remains constrained.

%%%%%%%%%%%%%%%%%%%%%%%%%%%%%%%%%%%%%%%%
%%%%%%%%%%%%%%%%%%%%%%%%%%%%%%%%%%%%%%%%
\begin{figure*}[htb!]
\centering
\includegraphics[width=\textwidth]{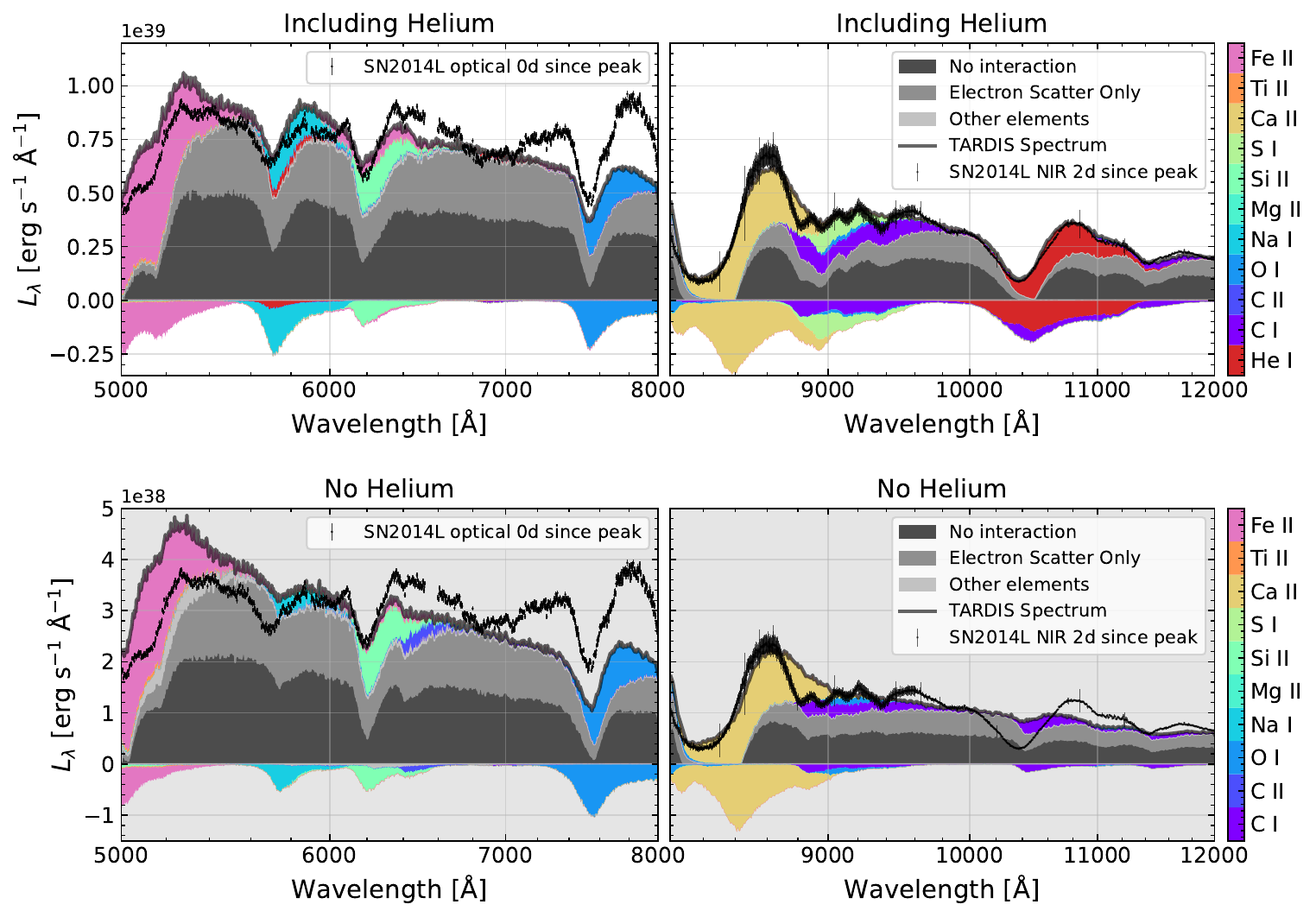}
\caption{The SDEC plot shows the spectral-energy decomposition for the \tardis spectrum evaluated using the MAP parameter set (top row) compared with an inference run in which we remove He (bottom row). 
The optical region (left column) remains nearly identical between the two models, with no major differences in the strong line profiles. 
In contrast, the NIR region (right column) demonstrates that the 1~\um feature cannot be reproduced without He. 
For reference, we also plot the observed spectra; however, these are not continuum-matched to the synthetic spectra.}
\label{fig: SDEC_comparison}
\end{figure*}
%%%%%%%%%%%%%%%%%%%%%%%%%%%%%%%%%%%%%%%%
%%%%%%%%%%%%%%%%%%%%%%%%%%%%%%%%%%%%%%%%

%%%%%%%%%%%%%%%%%%%%%%%%%%%%%%%%%%%%%%%
\subsubsection{Helium}
%%%%% Inferred He masses
Full spectral modeling of the optical and NIR spectra near peak light indicates a statistically significant, non-zero He component ($0.019^{0.020}_{0.018}$~\Msun) above \vinner $=\vinnervalue$~\kms in SN~2014L. 
A comparison inference run with no He fails to reproduce key observed features, particularly the pronounced $1~\mu$m absorption in the NIR, demonstrating that some He must be present (see the lower panels of Figure~\ref{fig: SDEC_comparison}). 
The inferred He mass lies below previously reported upper limits for He-poor SNe \citep[\eg][]{hachinger_how_2012, williamson_modeling_2021, shahbandeh_carnegie_2022, kumar_near-infrared_2025}.

%%%%% No He lines in the optical 
In the top panels of Figure~\ref{fig: SDEC_comparison}, we compare the observed spectra with the \tardis simulation and show the corresponding elemental energy decomposition. 
We note that the synthetic and observed spectra are not continuum-matched in this comparison. 
In the optical region, the \tardis spectrum exhibits no distinct \ion{He}{1} features. 
The elemental-decomposition plot indicates that the P-Cygni feature near $\lambda 5500$~\AA\ contains a small \ion{He}{1} contribution, but this component is heavily dominated by \ion{Na}{1}~D.

%%%% NIR 1um feature
\citet{shahbandeh_carnegie_2022} compared the observed NIR spectra of SN~2014L with the model spectra of \citet{teffs_type_2020} and suggested that the strong absorption feature near 1~\um may arise from a mixture of a small portion of \ion{He}{1} together with \ion{C}{1}, \ion{S}{1}, and/or \ion{Mg}{2}. 
From our inference, the \tardis simulation of the inferred MAP parameter set indicates the $1~\mu$m absorption feature observed in SN~2014L is dominated by \ion{He}{1} 1.083~\um triplets ($\sim$70\% pacaket contribution in wavelength region between 1.00~\um and 1.07~\um) and partially blended with \ion{C}{1} ($\sim$30\% packet contribution).

%%%% He level populations
Our model also reproduces the weak \ion{He}{1} 2.058~\um feature in the observed spectrum, see Figure~\ref{fig: posterior_spectral_comparison}.
The difference in spectral strength between the \ion{He}{1} 1.038~\um\ and 2.058~\um\ lines is not expected to follow a fixed ratio \citep[\eg,][]{patat_metamorphosis_2001}. Observational study of SNe~Ib NIR spectra shows that the strength ratio between the absorption features at 1~\um\ and 2~\um\ ranges from 5 to 15 \citep{shahbandeh_carnegie_2022}.
In our MAP \tardis simulation, the level population responsible for the lower level of the \ion{He}{1} 1.083~\um line transition ($2^3\textrm{S}$) is a factor of 10 more than the lower level of the 2.058~\um line ($2^1\textrm{S}$), which explains the strong 1.083~\um but weak 2.058~\um feature. 
In the comparison inference run that has no He in the prior, neither the strong $1~\mu$m feature (see the bottom right panel of Figure~\ref{fig: SDEC_comparison}) nor the weak $2~\mu$m feature is reproduced. 

%%%%% Significance of this work 
Given the strong agreement between the observed and inferred spectra of SN~2014L, a representative SN~Ic, our results suggest that the $1~\mu$m feature in at least some SNe~Ic is likely dominated by He. 
This conclusion is consistent with the broader interpretation of \citet{shahbandeh_carnegie_2022}, which concluded that residual He may be more common than previously assumed and highlighted the essential role of NIR spectra in constraining He in SNe~Ic.

%%%%%%%%%%%%%%%%%%%%%%%%%%%%%%%%%%%%%%%
\subsubsection{Carbon and Oxygen}
%%%%% C I features 
We infer $1.25^{1.27}_{1.22}$~\Msun of C above \vinner = $\vinnervalue$~\kms for SN~2014L.
In the \tardis simulation using the MAP parameter set, the majority of the C contribution in the spectra arises from \ion{C}{1} transitions in the NIR. 
As shown in the elemental-decomposition plots in Figures~\ref{fig: posterior_spectral_comparison} and \ref{fig: SDEC_comparison}, there is siginificant \ion{C}{1} contribution around 0.9~\um, blended with \ion{S}{1}.
The model does not fully reproduce the distinct absorption minima in this region, likely due to the assumption of a uniform abundance structure, which has a stronger impact at redder wavelengths. 
The absorption features near 1.15, 1,40 and 1.70~\um are attributed to \ion{C}{1} 1.1754~\um, 1.4543~\um and 1.7340~\um line transitions, respectively. 
We note that in the observed spectrum, the latter three features are affected by telluric absorptions.

%%%%% C-rich outer ejecta
Excluding Ne, C is the most abundant element in the outer ejecta, indicating that SN~2014L has a C-rich outer envelope. 
Previous studies have associated the presence of \ion{C}{1} absorption features in SNe~Ic, such as SN~1994I \citep{baron_preliminary_1996}, SN~2007gr \citep{valenti_carbon-rich_2008}, and SN~2014L \citep{zhang_optical_2018}, with C-rich progenitors. 
However, the C features in SN~2014L exhibit strengths comparable to those of typical SNe~Ic \citep{zhang_optical_2018}, and are different from the peculiar 2019ewu-like C-rich SNe with strong \ion{C}{2} feature \citep{williamson_sn_2023}. 

%%%%%% O I feature
The inference results indicate $0.10^{0.11}_{0.09}$~\Msun\ of O above \vinner. 
Within the simulated wavelength range, the only clearly identifiable O feature in the synthetic spectra is the \ion{O}{1} $\lambda$7774~\AA\ line in the optical. 
The \ion{O}{1} 0.9264~\um transition contributes only at the $\sim$5\% level to the absorption feature near 0.9~\um, which is dominated by \ion{C}{1} and \ion{S}{1}, as discussed above.

%%%%%% High C/O ratio
These results potentially suggest a high C/O ratio of $13^{14}_{12}$ in the outer ejecta of SN~2014L. 
If the composition of the outer envelope responsible for the peak emission does not change substantially from the pre-explosion surface layers \citep[\eg][]{laplace_different_2021}, then the progenitor of SN~2014L likely possessed a high surface C/O ratio. 
We emphasize that the surface abundance does not represent the core composition or the total nucleosynthetic yield, as the surface layers are highly sensitive to mixing and mass-loss processes \citep[\eg][]{georgy_grids_2012, groh_grids_2019, Ekstrom2021}. 
Stellar-evolution calculations show that high surface C/O ratios, together with low surface He masses, can arise in models with strong mass loss, whether through winds or binary interaction, particularly for higher-mass progenitors \citep[\eg][]{yoon_type_2010, ma_carbon_2025}.

%%%%%%%%%%%%%%%%%%%%%%%%%%%%%%%%%%%%%%%
\subsubsection{Intermediate Mass Elements}
%%%%% Report IME elemental mass above the v_inner
Excluding Ne (which remains unconstrained), we infer a total IME mass of $0.46^{0.50}_{0.42}$~\Msun\ above \vinner. 
Within this group, S attains the highest mass fraction and is inferred to be the second-most abundant element in the outer ejecta after C. 
However, varying the S mass while holding the other elemental masses fixed at their MAP values produces only minor changes in the synthetic spectra. 
Previous modeling studies of SNe~Ib/c typically find S masses that are comparable to or lower than those of the other IMEs in the outer ejecta layers \citep[\eg][]{hachinger_how_2012, frey_can_2013, teffs_observations_2021}.

%%%%% Na and Si
Na and Si have comparable inferred masses above \vinner, each on the order of $10^{-2}$~\Msun. 
The elemental-decomposition plot from the \tardis simulation shows that \ion{Na}{1}~D ($\lambda\lambda$5890, 5896~\AA) dominates the line interactions producing the absorption near $\lambda$5600~\AA, with only a minor contribution from \ion{He}{1}~$\lambda$5876~\AA.

%%%%% Mg 
Mg is the least abundant IME in our inference, with only a few $10^{-5}$~\Msun\ above \vinner. 
The reconstructed spectra show no distinct Mg features, as illustrated in the bottom panel of Figure~\ref{fig: posterior_spectral_comparison} and in the top panels of Figure~\ref{fig: SDEC_comparison}.
Although previous spectral studies on SNe~Ic suggested that the strong absorption feature near 1~\um contains Mg \citep[\eg,][]{williamson_modeling_2021, shahbandeh_carnegie_2022}, we do not see Mg contribution to this feature in SN~2014L.

%%%%% Ca results
The top panel of Figure~\ref{fig: posterior_spectral_comparison} shows that our inferred model reproduces both the \ion{Ca}{2} H\&K features in the near-UV and the \ion{Ca}{2} NIR triplet with precision. 
We infer Ca mass above \vinner is on the order of $10^{-4}$~\Msun.
Initial experiment with a uniform Ca resulted in strong high-velocity features, which were inconsistent with the observed Ca features. 
Thus, we imposed a transition velocity \vzeroCa, above which there is no Ca (see Section~\ref{subsec: method - SN model}). 
\vzeroCa has an inferred value of $18221^{18281}_{18162}$~\kms.
This modified Ca distribution aligns with the prescription of \citet{hachinger_how_2012}, who introduced outer layers with reduced IME abundances, particularly Ca, to suppress high-velocity Ca features which are usually not seen in SN~Ic spectra.

\begin{figure}[tbh!]
\centering
\includegraphics[width=\columnwidth]{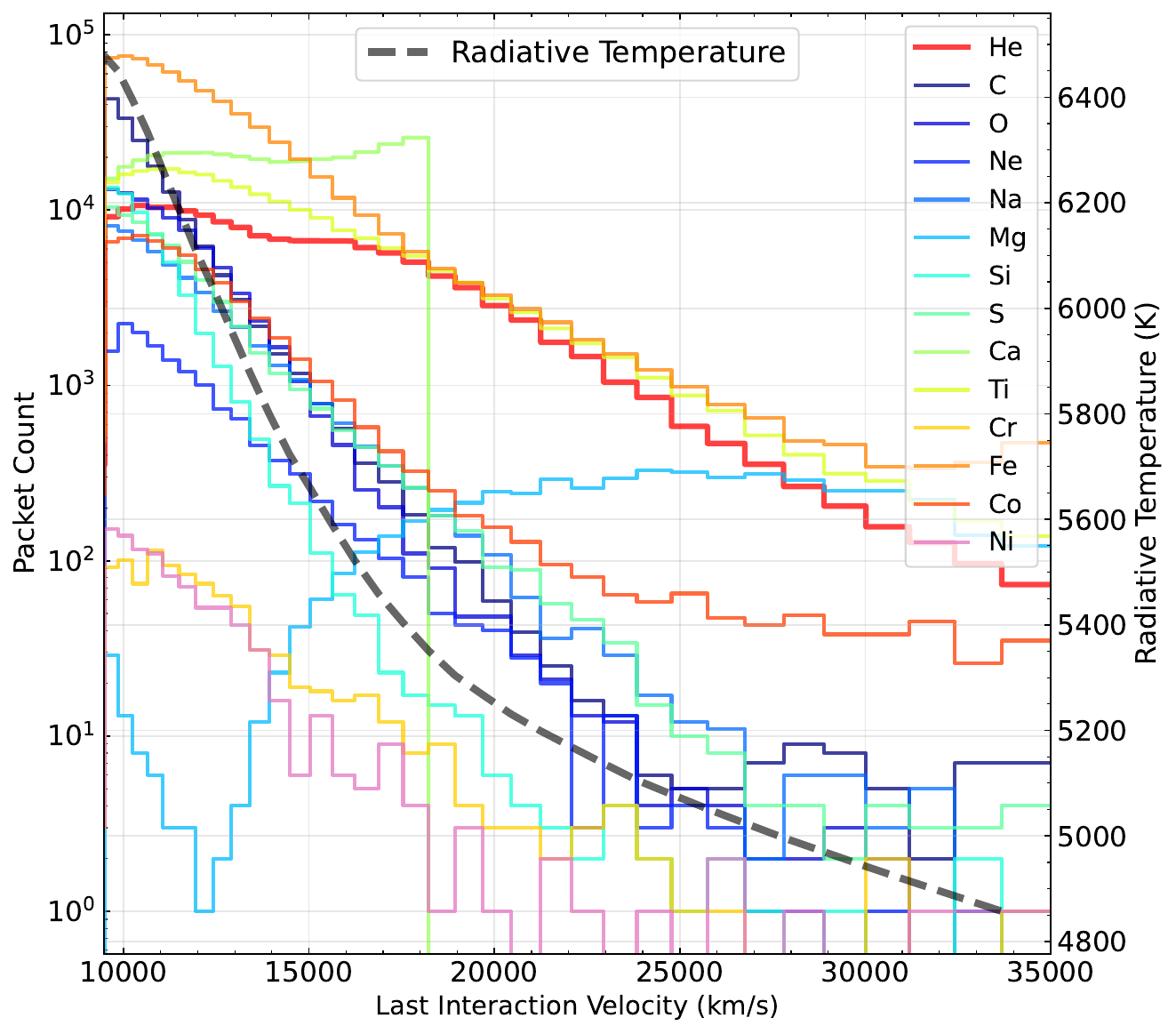}
\caption{The last line interaction velocity distribution of the \tardis simulation ran with the MAP parameter set. Each line represents the velocity distribution of the packets that line interacted with a specific element as their last interaction before crossing the outermost simulation boundary.}
\label{fig: LIV_comparison}
\end{figure}

%%%%% Ca line interaction distribution
Figure~\ref{fig: LIV_comparison} shows the distribution of last line-interaction locations from the \tardis simulation as a function of ejecta velocity. 
In our modified uniform-abundance ejecta model, the line-interaction probability of Ca is nearly uniform throughout all regions where it is present, whereas other elements interact predominantly near \vinner. 
This behavior arises because \ion{Ca}{2} remains abundant even in the low-density outer layers, where its comparatively low ionization/excitation energies allow line interactions to occur at relatively low radiative temperatures. 
The resulting extended spatial distribution of \ion{Ca}{2} interactions can therefore account for the higher observed line velocities \citep{zhang_optical_2018}, without compositional enhancement of Ca at large radii.

%%%%% Ca strong is not the same as Ca rich 
By comparing with other observed SNe~Ic, \citet{zhang_optical_2018} suggests a potential negative correlation between the Ca mass and the SN luminosity, based on measurements of the \ion{Ca}{2} NIR triplet feature strength. 
However, Ca has strong lines that are easily saturated. 
The \ion{Ca}{2} NIR triplet profile is especially more sensitive to radiative temperature than \ion{Ca}{2} H\&K. 
Given the same composition above \vinner, a model with higher input luminosity, which leads to higher radiative temperature, can produce a weaker \ion{Ca}{2} NIR triplet profile due to Ca atoms being populated at a higher ionization/excitation states than the ones required for the $\lambda\lambda$8498, 8542, 8662~\AA\ line transitions. 

%%%%%%%%%%%%%%%%%%%%%%%%%%%%%%%%%%%%%%%%
%%%%%%%%%%%%%%%%%%%%%%%%%%%%%%%%%%%%%%%%
%%%%%%%%%%%%%%%%%%%%%%%%%%%%%%%%%%%%%%%%
%%%%%%%%%%%%%%%%%%%%%%%%%%%%%%%%%%%%%%%%

%%%%%%%%%%%%%%%%%%%%%%%%%%%%%%%%%%%%%%%
\subsubsection{Iron Group Elements}
%%%%%% IGE results
SN~2014L is inferred to contain $0.0028^{0.0039}_{0.0026}$~\Msun of IGEs above \vinner, which contributes the least in terms of mass compared to other elements.
Note that the inferred IGE mass corresponds to the existing amount near peak time, not the raw amount before the radioactive decay takes effect. 

%%%%%% IGE features
The most prominent IGE feature in SN~2014L is \ion{Fe}{2}, and is broadly reproduced in the MAP spectra, see Figure~\ref{fig: posterior_spectral_comparison}.
Minor mismatches remain in several weaker features, potentially reflecting contributions from elements not included in our model and/or limitations of the uniform-abundance assumption. 
Additional observation in the UV, where many IGE transitions are concentrated, would help refine the IGE mass inference.

%%%%%%%%%%%%%%%%%%%%%%%%%%%%%%%%%%%%%%%%%%%%%%%%%%%%%%%%%%%%%%%%%%%%%%%%%%%%%%%%
\section{Conclusions} \label{sec:conclusion}

%%%%% Summary of what we did in this work 
This work models the outer ejecta of an SN~Ic 2014L near peak light using one optical spectrum from \citet{zhang_optical_2018} and one NIR spectrum from \citet{shahbandeh_carnegie_2022}. 
We apply a Bayesian inference framework that emulates \tardis radiative transfer calculations with a probabilistic deep-learning model \citep{obrien_probabilistic_2021, obrien_1991t-like_2023, kerzendorf_probabilistic_2022}. 
This approach enables a quantitative determination of the outer ejecta composition and density structure of SN~2014L, together with their uncertainties and posterior correlations.

%%%%%% He detection 
SN2014L requires a small but non-zero amount of helium in the outer ejecta to have a model consistent with the observed spectra.
An inference identical in all respects but lacking He fails to match the observed strong features in the NIR, even though the optical features remain similar. 
The absorption feature near $1~\mu$m is reproduced only when He is included, and the \tardis spectral-decomposition output shows that the feature is dominated by \ion{He}{1} using the maximum-a-posteriori parameter set. 
The inferred He mass above the photosphere (\vinner$=\vinnervalue$~\kms) at peak light is small (0.018 to 0.020~\Msun), consistent with the upper limits inferred for He-poor SNe \citep[\eg][]{hachinger_how_2012, dessart_radiative-transfer_2015, williamson_modeling_2021, shahbandeh_carnegie_2022, kumar_near-infrared_2025}.

%%%%% Density 
We infer a power-law density exponent of $-7.04$ to $-6.88$ (16\% to 84\% credible interval) for the outer ejecta of SN~2014L near maximum light, consistent with the radiation-dominated explosions in theoretical calculations \citep[\eg,][]{colgate_early_1969,chevalier_hydrodynamics_1976,chevalier_exploding_1981,iwamoto_theoretical_1994}.

%%%%% Composition
The inferred composition indicates a C-rich outer ejecta with a high C/O mass ratio.
Despite strong observed \ion{Ca}{2} features, the Ca mass above the photosphere is a few $10^{-4}$~\Msun, showing that a strong Ca feature does not necessarily imply a high Ca abundance. 
The posterior favors a stratified Ca profile with \vzeroCa of 18162 to 18281~\kms, consistent with literature prescriptions that suppress excessive high-velocity Ca \citep[\eg,][]{hachinger_how_2012}.

%%%%% Outlook
This study demonstrates that a Bayesian radiative-transfer inference framework accelerated with a probabilistic emulator provides a rigorous and computationally feasible means to quantify ejecta composition. 
Future work will need to apply methodology to larger optical and NIR samples, including SNe~Ib that show both strong 1~\um and 2~\um He features \citep[\eg,][]{shahbandeh_carnegie_2022, modjaz_shock_2009}, to enable systematic constraints on residual He, C/O ratios, and envelope stripping pathways in SNe~Ib/Ic progenitors.

%%%%%%%%%%%%%%%%%%%%%%%%%%%%%%%%%%%%%%%%%%%%%%%%%%%%%%%%%%%%%%%%%%%%%%%%%%%%%%%%
%%%%%%%%%%%%%%%%%%%%%%%%%%%%%%%%%%%%%%%%%%%%%%%%%%%%%%%%%%%%%%%%%%%%%%%%%%%%%%%%
\bibliography{SN_Ref, software_ref, tardis_zotero}

\begin{thebibliography}{}
\expandafter\ifx\csname natexlab\endcsname\relax\def\natexlab#1{#1}\fi
\providecommand{\url}[1]{\href{#1}{#1}}
\providecommand{\dodoi}[1]{doi:~\href{http://doi.org/#1}{\nolinkurl{#1}}}
\providecommand{\doeprint}[1]{\href{http://ascl.net/#1}{\nolinkurl{http://ascl.net/#1}}}
\providecommand{\doarXiv}[1]{\href{https://arxiv.org/abs/#1}{\nolinkurl{https://arxiv.org/abs/#1}}}

\bibitem[{C. Ashall \& P.~A. Mazzali(2020)Ashall \& Mazzali}]{ashall_extracting_2020}
Ashall, C., \& Mazzali, P.~A. 2020, \bibinfo{title}{Extracting high-level information from gamma-ray burst supernova spectra,} Monthly Notices of the Royal Astronomical Society, 492, 5956, \dodoi{10.1093/mnras/staa212}

\bibitem[{G. Ashton {et~al.}(2022)Ashton, Bernstein, Buchner, Chen, Csányi, Fowlie, Feroz, Griffiths, Handley, Habeck, Higson, Hobson, Lasenby, Parkinson, Pártay, Pitkin, Schneider, Speagle, South, Veitch, Wacker, Wales, \& Yallup}]{ashton_nested_2022}
Ashton, G., Bernstein, N., Buchner, J., {et~al.} 2022, \bibinfo{title}{Nested sampling for physical scientists,} Nature Reviews Methods Primers, 2, 39, \dodoi{10.1038/s43586-022-00121-x}

\bibitem[{ {Astropy Collaboration} {et~al.}(2013){Astropy Collaboration}, {Robitaille}, {Tollerud}, {Greenfield}, {Droettboom}, {Bray}, {Aldcroft}, {Davis}, {Ginsburg}, {Price-Whelan}, {Kerzendorf}, {Conley}, {Crighton}, {Barbary}, {Muna}, {Ferguson}, {Grollier}, {Parikh}, {Nair}, {Unther}, {Deil}, {Woillez}, {Conseil}, {Kramer}, {Turner}, {Singer}, {Fox}, {Weaver}, {Zabalza}, {Edwards}, {Azalee Bostroem}, {Burke}, {Casey}, {Crawford}, {Dencheva}, {Ely}, {Jenness}, {Labrie}, {Lim}, {Pierfederici}, {Pontzen}, {Ptak}, {Refsdal}, {Servillat}, \& {Streicher}}]{astropy:2013}
{Astropy Collaboration}, {Robitaille}, T.~P., {Tollerud}, E.~J., {et~al.} 2013, \bibinfo{title}{{Astropy: A community Python package for astronomy},} \aap, 558, A33, \dodoi{10.1051/0004-6361/201322068}

\bibitem[{ {Astropy Collaboration} {et~al.}(2018){Astropy Collaboration}, {Price-Whelan}, {Sip{\H{o}}cz}, {G{\"u}nther}, {Lim}, {Crawford}, {Conseil}, {Shupe}, {Craig}, {Dencheva}, {Ginsburg}, {Vand erPlas}, {Bradley}, {P{\'e}rez-Su{\'a}rez}, {de Val-Borro}, {Aldcroft}, {Cruz}, {Robitaille}, {Tollerud}, {Ardelean}, {Babej}, {Bach}, {Bachetti}, {Bakanov}, {Bamford}, {Barentsen}, {Barmby}, {Baumbach}, {Berry}, {Biscani}, {Boquien}, {Bostroem}, {Bouma}, {Brammer}, {Bray}, {Breytenbach}, {Buddelmeijer}, {Burke}, {Calderone}, {Cano Rodr{\'\i}guez}, {Cara}, {Cardoso}, {Cheedella}, {Copin}, {Corrales}, {Crichton}, {D'Avella}, {Deil}, {Depagne}, {Dietrich}, {Donath}, {Droettboom}, {Earl}, {Erben}, {Fabbro}, {Ferreira}, {Finethy}, {Fox}, {Garrison}, {Gibbons}, {Goldstein}, {Gommers}, {Greco}, {Greenfield}, {Groener}, {Grollier}, {Hagen}, {Hirst}, {Homeier}, {Horton}, {Hosseinzadeh}, {Hu}, {Hunkeler}, {Ivezi{\'c}}, {Jain}, {Jenness}, {Kanarek}, {Kendrew}, {Kern}, {Kerzendorf}, {Khvalko}, {King}, {Kirkby}, {Kulkarni},
  {Kumar}, {Lee}, {Lenz}, {Littlefair}, {Ma}, {Macleod}, {Mastropietro}, {McCully}, {Montagnac}, {Morris}, {Mueller}, {Mumford}, {Muna}, {Murphy}, {Nelson}, {Nguyen}, {Ninan}, {N{\"o}the}, {Ogaz}, {Oh}, {Parejko}, {Parley}, {Pascual}, {Patil}, {Patil}, {Plunkett}, {Prochaska}, {Rastogi}, {Reddy Janga}, {Sabater}, {Sakurikar}, {Seifert}, {Sherbert}, {Sherwood-Taylor}, {Shih}, {Sick}, {Silbiger}, {Singanamalla}, {Singer}, {Sladen}, {Sooley}, {Sornarajah}, {Streicher}, {Teuben}, {Thomas}, {Tremblay}, {Turner}, {Terr{\'o}n}, {van Kerkwijk}, {de la Vega}, {Watkins}, {Weaver}, {Whitmore}, {Woillez}, {Zabalza}, \& {Astropy Contributors}}]{astropy:2018}
{Astropy Collaboration}, {Price-Whelan}, A.~M., {Sip{\H{o}}cz}, B.~M., {et~al.} 2018, \bibinfo{title}{{The Astropy Project: Building an Open-science Project and Status of the v2.0 Core Package},} \aj, 156, 123, \dodoi{10.3847/1538-3881/aabc4f}

\bibitem[{ {Astropy Collaboration} {et~al.}(2022){Astropy Collaboration}, {Price-Whelan}, {Lim}, {Earl}, {Starkman}, {Bradley}, {Shupe}, {Patil}, {Corrales}, {Brasseur}, {N{\"o}the}, {Donath}, {Tollerud}, {Morris}, {Ginsburg}, {Vaher}, {Weaver}, {Tocknell}, {Jamieson}, {van Kerkwijk}, {Robitaille}, {Merry}, {Bachetti}, {G{\"u}nther}, {Aldcroft}, {Alvarado-Montes}, {Archibald}, {B{\'o}di}, {Bapat}, {Barentsen}, {Baz{\'a}n}, {Biswas}, {Boquien}, {Burke}, {Cara}, {Cara}, {Conroy}, {Conseil}, {Craig}, {Cross}, {Cruz}, {D'Eugenio}, {Dencheva}, {Devillepoix}, {Dietrich}, {Eigenbrot}, {Erben}, {Ferreira}, {Foreman-Mackey}, {Fox}, {Freij}, {Garg}, {Geda}, {Glattly}, {Gondhalekar}, {Gordon}, {Grant}, {Greenfield}, {Groener}, {Guest}, {Gurovich}, {Handberg}, {Hart}, {Hatfield-Dodds}, {Homeier}, {Hosseinzadeh}, {Jenness}, {Jones}, {Joseph}, {Kalmbach}, {Karamehmetoglu}, {Ka{\l}uszy{\'n}ski}, {Kelley}, {Kern}, {Kerzendorf}, {Koch}, {Kulumani}, {Lee}, {Ly}, {Ma}, {MacBride}, {Maljaars}, {Muna}, {Murphy}, {Norman},
  {O'Steen}, {Oman}, {Pacifici}, {Pascual}, {Pascual-Granado}, {Patil}, {Perren}, {Pickering}, {Rastogi}, {Roulston}, {Ryan}, {Rykoff}, {Sabater}, {Sakurikar}, {Salgado}, {Sanghi}, {Saunders}, {Savchenko}, {Schwardt}, {Seifert-Eckert}, {Shih}, {Jain}, {Shukla}, {Sick}, {Simpson}, {Singanamalla}, {Singer}, {Singhal}, {Sinha}, {Sip{\H{o}}cz}, {Spitler}, {Stansby}, {Streicher}, {{\v{S}}umak}, {Swinbank}, {Taranu}, {Tewary}, {Tremblay}, {de Val-Borro}, {Van Kooten}, {Vasovi{\'c}}, {Verma}, {de Miranda Cardoso}, {Williams}, {Wilson}, {Winkel}, {Wood-Vasey}, {Xue}, {Yoachim}, {Zhang}, {Zonca}, \& {Astropy Project Contributors}}]{astropy:2022}
{Astropy Collaboration}, {Price-Whelan}, A.~M., {Lim}, P.~L., {et~al.} 2022, \bibinfo{title}{{The Astropy Project: Sustaining and Growing a Community-oriented Open-source Project and the Latest Major Release (v5.0) of the Core Package},} \apj, 935, 167, \dodoi{10.3847/1538-4357/ac7c74}

\bibitem[{E. Baron {et~al.}(1999)Baron, Branch, Hauschildt, Filippenko, \& Kirshner}]{baron_spectral_1999}
Baron, E., Branch, D., Hauschildt, P.~H., Filippenko, A.~V., \& Kirshner, R.~P. 1999, \bibinfo{title}{Spectral {Models} of the {Type} {IC} {Supernova} {SN} {1994I} in {M51},} The Astrophysical Journal, 527, 739, \dodoi{10.1086/308107}

\bibitem[{E. Baron {et~al.}(1996)Baron, Hauschildt, Branch, Kirshner, \& Filippenko}]{baron_preliminary_1996}
Baron, E., Hauschildt, P.~H., Branch, D., Kirshner, R.~P., \& Filippenko, A.~V. 1996, \bibinfo{title}{Preliminary spectral analysis of {SN} {1994I},} Monthly Notices of the Royal Astronomical Society, 279, 799, \dodoi{10.1093/mnras/279.3.799}

\bibitem[{A. Boyle {et~al.}(2017)Boyle, Sim, Hachinger, \& Kerzendorf}]{boyle_helium_2017}
Boyle, A., Sim, S.~A., Hachinger, S., \& Kerzendorf, W. 2017, \bibinfo{title}{Helium in double-detonation models of type {Ia} supernovae,} Astronomy and Astrophysics, 599, A46, \dodoi{10.1051/0004-6361/201629712}

\bibitem[{J. Buchner(2021)Buchner}]{buchner_ultranest_2021}
Buchner, J. 2021, \bibinfo{title}{{UltraNest} - a robust, general purpose {Bayesian} inference engine,} Journal of Open Source Software, 6, 3001, \dodoi{10.21105/joss.03001}

\bibitem[{R.~A. Chevalier(1976)Chevalier}]{chevalier_hydrodynamics_1976}
Chevalier, R.~A. 1976, \bibinfo{title}{The hydrodynamics of type {II} supernovae.,} The Astrophysical Journal, 207, 872, \dodoi{10.1086/154557}

\bibitem[{R.~A. Chevalier(1981)Chevalier}]{chevalier_exploding_1981}
Chevalier, R.~A. 1981, \bibinfo{title}{Exploding white dwarf models for type {I} {SN}.,} The Astrophysical Journal, 246, 267, \dodoi{10.1086/158920}

\bibitem[{A. Clocchiatti {et~al.}(1996)Clocchiatti, Wheeler, Brotherton, Cochran, Wills, Barker, \& Turatto}]{clocchiatti_sn_1996}
Clocchiatti, A., Wheeler, J.~C., Brotherton, M.~S., {et~al.} 1996, \bibinfo{title}{{SN} {1994I}: {Disentangling} {He} i {Lines} in {Type} {IC} {Supernovae},} The Astrophysical Journal, 462, 462, \dodoi{10.1086/177165}

\bibitem[{S.~A. Colgate \& C. McKee(1969)Colgate \& McKee}]{colgate_early_1969}
Colgate, S.~A., \& McKee, C. 1969, \bibinfo{title}{Early {Supernova} {Luminosity},} The Astrophysical Journal, 157, 623, \dodoi{10.1086/150102}

\bibitem[{P.~A. Crowther(2007)Crowther}]{crowther_physical_2007}
Crowther, P.~A. 2007, \bibinfo{title}{Physical {Properties} of {Wolf}-{Rayet} {Stars},} Annual Review of Astronomy and Astrophysics, 45, 177, \dodoi{10.1146/annurev.astro.45.051806.110615}

\bibitem[{L. Dessart \& D.~J. Hillier(2015)Dessart \& Hillier}]{dessart_one-dimensional_2015}
Dessart, L., \& Hillier, D.~J. 2015, \bibinfo{title}{One-dimensional non-{LTE} time-dependent radiative transfer of an {He}-detonation model and the connection to faint and fast-decaying supernovae,} Monthly Notices of the Royal Astronomical Society, 447, 1370, \dodoi{10.1093/mnras/stu2520}

\bibitem[{L. Dessart {et~al.}(2012)Dessart, Hillier, Li, \& Woosley}]{dessart_nature_2012}
Dessart, L., Hillier, D.~J., Li, C., \& Woosley, S. 2012, \bibinfo{title}{On the nature of supernovae {Ib} and {Ic},} Monthly Notices of the Royal Astronomical Society, 424, 2139, \dodoi{10.1111/j.1365-2966.2012.21374.x}

\bibitem[{L. Dessart {et~al.}(2011)Dessart, Hillier, Livne, Yoon, Woosley, Waldman, \& Langer}]{dessart_core-collapse_2011}
Dessart, L., Hillier, D.~J., Livne, E., {et~al.} 2011, \bibinfo{title}{Core-collapse explosions of {Wolf}-{Rayet} stars and the connection to {Type} {IIb}/{Ib}/{Ic} supernovae,} Monthly Notices of the Royal Astronomical Society, 414, 2985, \dodoi{10.1111/j.1365-2966.2011.18598.x}

\bibitem[{L. Dessart {et~al.}(2015)Dessart, Hillier, Woosley, Livne, Waldman, Yoon, \& Langer}]{dessart_radiative-transfer_2015}
Dessart, L., Hillier, D.~J., Woosley, S., {et~al.} 2015, \bibinfo{title}{Radiative-transfer models for supernovae {IIb}/{Ib}/{Ic} from binary-star progenitors,} Monthly Notices of the Royal Astronomical Society, 453, 2189, \dodoi{10.1093/mnras/stv1747}

\bibitem[{L. Dessart {et~al.}(2020)Dessart, Yoon, Aguilera-Dena, \& Langer}]{dessart_supernovae_2020}
Dessart, L., Yoon, S.-C., Aguilera-Dena, D.~R., \& Langer, N. 2020, \bibinfo{title}{Supernovae {Ib} and {Ic} from the explosion of helium stars,} Astronomy and Astrophysics, 642, A106, \dodoi{10.1051/0004-6361/202038763}

\bibitem[{C. Dugas {et~al.}(2000)Dugas, Bengio, B{\'e}lisle, Nadeau, \& Garcia}]{dugas2000incorporating}
Dugas, C., Bengio, Y., B{\'e}lisle, F., Nadeau, C., \& Garcia, R. 2000, \bibinfo{title}{Incorporating second-order functional knowledge for better option pricing,} Advances in neural information processing systems, 13

\bibitem[{S. {Ekstr{\"o}m}(2021){Ekstr{\"o}m}}]{Ekstrom2021}
{Ekstr{\"o}m}, S. 2021, \bibinfo{title}{{Massive star modelling and nucleosynthesis},} Frontiers in Astronomy and Space Sciences, 8, 53, \dodoi{10.3389/fspas.2021.617765}

\bibitem[{J.~J. Eldridge {et~al.}(2013)Eldridge, Fraser, Smartt, Maund, \& Crockett}]{eldridge_death_2013}
Eldridge, J.~J., Fraser, M., Smartt, S.~J., Maund, J.~R., \& Crockett, R.~M. 2013, \bibinfo{title}{The death of massive stars - {II}. {Observational} constraints on the progenitors of {Type} {Ibc} supernovae,} Monthly Notices of the Royal Astronomical Society, 436, 774, \dodoi{10.1093/mnras/stt1612}

\bibitem[{A. Ercolino {et~al.}(2023)Ercolino, Jin, Langer, \& Dessart}]{ercolino_interacting_2023}
Ercolino, A., Jin, H., Langer, N., \& Dessart, L. 2023, \bibinfo{title}{Interacting supernovae from wide massive binary systems,} \dodoi{10.48550/arXiv.2308.01819}

\bibitem[{W. Falcon \& T.~P.~L. team(2024)Falcon \& team}]{falcon_pytorch_2024}
Falcon, W., \& team, T. P.~L. 2024, \bibinfo{title}{{PyTorch} {Lightning},} Zenodo, \dodoi{10.5281/zenodo.13254264}

\bibitem[{Q. Fang {et~al.}(2019)Fang, Maeda, Kuncarayakti, Sun, \& Gal-Yam}]{fang_hybrid_2019}
Fang, Q., Maeda, K., Kuncarayakti, H., Sun, F., \& Gal-Yam, A. 2019, \bibinfo{title}{A hybrid envelope-stripping mechanism for massive stars from supernova nebular spectroscopy,} Nature Astronomy, 3, 434, \dodoi{10.1038/s41550-019-0710-6}

\bibitem[{R. Farmer {et~al.}(2023)Farmer, Laplace, Ma, de~Mink, \& Justham}]{farmer_nucleosynthesis_2023}
Farmer, R., Laplace, E., Ma, J.-z., de~Mink, S.~E., \& Justham, S. 2023, \bibinfo{title}{Nucleosynthesis of {Binary}-stripped {Stars},} The Astrophysical Journal, 948, 111, \dodoi{10.3847/1538-4357/acc315}

\bibitem[{A.~V. Filippenko(1997)Filippenko}]{filippenko_optical_1997}
Filippenko, A.~V. 1997, \bibinfo{title}{Optical {Spectra} of {Supernovae},} {\textbackslash}araa, 35, 309, \dodoi{10.1146/annurev.astro.35.1.309}

\bibitem[{A.~V. Filippenko {et~al.}(1995)Filippenko, Barth, Matheson, Armus, Brown, Espey, Fan, Goodrich, Ho, Junkkarinen, Koo, Lehnert, Martel, Mazzarella, Miller, Smith, Tytler, \& Wirth}]{filippenko_type_1995}
Filippenko, A.~V., Barth, A.~J., Matheson, T., {et~al.} 1995, \bibinfo{title}{The {Type} {IC} {Supernova} {1994I} in {M51}: {Detection} of {Helium} and {Spectral} {Evolution},} The Astrophysical Journal, 450, L11, \dodoi{10.1086/309659}

\bibitem[{L.~H. Frey {et~al.}(2013)Frey, Fryer, \& Young}]{frey_can_2013}
Frey, L.~H., Fryer, C.~L., \& Young, P.~A. 2013, \bibinfo{title}{Can {Stellar} {Mixing} {Explain} the {Lack} of {Type} {Ib} {Supernovae} in {Long}-duration {Gamma}-{Ray} {Bursts}?} The Astrophysical Journal, 773, L7, \dodoi{10.1088/2041-8205/773/1/L7}

\bibitem[{A.~G. Fullard {et~al.}(2022)Fullard, O'Brien, Kerzendorf, Shrestha, Hoffman, Ignace, \& van~der Smagt}]{fullard_new_2022}
Fullard, A.~G., O'Brien, J.~T., Kerzendorf, W.~E., {et~al.} 2022, \bibinfo{title}{New {Mass} {Estimates} for {Massive} {Binary} {Systems}: {A} {Probabilistic} {Approach} {Using} {Polarimetric} {Radiative} {Transfer},} The Astrophysical Journal, 930, 89, \dodoi{10.3847/1538-4357/ac589e}

\bibitem[{A. Gal-Yam(2017)Gal-Yam}]{gal-yam_observational_2017}
Gal-Yam, A. 2017, in Handbook of {Supernovae}, 195, \dodoi{10.1007/978-3-319-21846-5_35}

\bibitem[{C. Georgy {et~al.}(2012)Georgy, Ekström, Meynet, Massey, Levesque, Hirschi, Eggenberger, \& Maeder}]{georgy_grids_2012}
Georgy, C., Ekström, S., Meynet, G., {et~al.} 2012, \bibinfo{title}{Grids of stellar models with rotation. {II}. {WR} populations and supernovae/{GRB} progenitors at {Z} = 0.014,} Astronomy and Astrophysics, 542, A29, \dodoi{10.1051/0004-6361/201118340}

\bibitem[{J.~H. Groh {et~al.}(2019)Groh, Ekström, Georgy, Meynet, Choplin, Eggenberger, Hirschi, Maeder, Murphy, Boian, \& Farrell}]{groh_grids_2019}
Groh, J.~H., Ekström, S., Georgy, C., {et~al.} 2019, \bibinfo{title}{Grids of stellar models with rotation. {IV}. {Models} from 1.7 to 120 {M}⊙ at a metallicity {Z} = 0.0004,} Astronomy and Astrophysics, 627, A24, \dodoi{10.1051/0004-6361/201833720}

\bibitem[{S. Hachinger(2011)Hachinger}]{hachinger_analysis_2011}
Hachinger, S. 2011, {PhD} {Thesis}, TU München

\bibitem[{S. Hachinger {et~al.}(2012)Hachinger, Mazzali, Taubenberger, Hillebrandt, Nomoto, \& Sauer}]{hachinger_how_2012}
Hachinger, S., Mazzali, P.~A., Taubenberger, S., {et~al.} 2012, \bibinfo{title}{How much {H} and {He} is 'hidden' in {SNe} {Ib}/c? - {I}. {Low}-mass objects,} Monthly Notices of the Royal Astronomical Society, 422, 70, \dodoi{10.1111/j.1365-2966.2012.20464.x}

\bibitem[{S. Hachinger {et~al.}(2017)Hachinger, Röpke, Mazzali, Gal-Yam, Maguire, Sullivan, Taubenberger, Ashall, Campbell, Elias-Rosa, Feindt, Greggio, Inserra, Miluzio, Smartt, \& Young}]{hachinger_type_2017}
Hachinger, S., Röpke, F.~K., Mazzali, P.~A., {et~al.} 2017, \bibinfo{title}{Type {Ia} supernovae with and without blueshifted narrow {Na} {I} {D} lines - how different is their structure?} {\textbackslash}mnras, 471, 491, \dodoi{10.1093/mnras/stx1578}

\bibitem[{R.~P. Harkness {et~al.}(1987)Harkness, Wheeler, Margon, Downes, Kirshner, Uomoto, Barker, Cochran, Dinerstein, Garnett, \& Levreault}]{harkness_early_1987}
Harkness, R.~P., Wheeler, J.~C., Margon, B., {et~al.} 1987, \bibinfo{title}{The {Early} {Spectral} {Phase} of {Type} {Ib} {Supernovae}: {Evidence} for {Helium},} The Astrophysical Journal, 317, 355, \dodoi{10.1086/165283}

\bibitem[{K. He {et~al.}(2016)He, Zhang, Ren, \& Sun}]{he_deep_2016}
He, K., Zhang, X., Ren, S., \& Sun, J. 2016, in 2016 {IEEE} {Conference} on {Computer} {Vision} and {Pattern} {Recognition} ({CVPR}), 770--778, \dodoi{10.1109/CVPR.2016.90}

\bibitem[{A. Heger {et~al.}(2003)Heger, Fryer, Woosley, Langer, \& Hartmann}]{heger_how_2003}
Heger, A., Fryer, C.~L., Woosley, S.~E., Langer, N., \& Hartmann, D.~H. 2003, \bibinfo{title}{How {Massive} {Single} {Stars} {End} {Their} {Life},} The Astrophysical Journal, 591, 288, \dodoi{10.1086/375341}

\bibitem[{E.~Y. {Hsiao} {et~al.}(2019){Hsiao}, {Philips}, {Marion}, {Kirshner}, {Morrell}, {Sand}, {Burns}, {Contreras}, {Hoeflich}, {Stritzinger}, {Valenti}, {Anderson}, {Ashall}, {Baltay}, {Baron}, {Banerjee}, {Davis}, {Diamond}, {Folatelli}, {Freedman}, {F{\"o}rster}, {Galbany}, {Gall}, {Gonz{\'a}lez-Gait{\'a}n}, {Goobar}, {Hamuy}, {Holmbo}, {Kasliwal}, {Krisciunas}, {Kumar}, {Lidman}, {Lu}, {Nugent}, {Perlmutter}, {Persson}, {Piro}, {Rabinowitz}, {Roth}, {Ryder}, {Schmidt}, {Shahbandeh}, {Suntzeff}, {Taddia}, {Uddin}, \& {Wang}}]{Hsiao2019}
{Hsiao}, E.~Y., {Philips}, M.~M., {Marion}, G.~H., {et~al.} 2019, \bibinfo{title}{{Carnegie Supernova Project-II: The Near-infrared Spectroscopy Program},} \pasp, 131, 014002, \dodoi{10.1088/1538-3873/aae961}

\bibitem[{G. Huang {et~al.}(2017)Huang, Liu, van~der Maaten, \& Weinberger}]{huang_densely_2017}
Huang, G., Liu, Z., van~der Maaten, L., \& Weinberger, K.~Q. 2017, 4700--4708.
\newblock \url{https://openaccess.thecvf.com/content_cvpr_2017/html/Huang_Densely_Connected_Convolutional_CVPR_2017_paper.html}

\bibitem[{D.~J. Hunter {et~al.}(2009)Hunter, Valenti, Kotak, Meikle, Taubenberger, Pastorello, Benetti, Stanishev, Smartt, Trundle, Arkharov, Bufano, Cappellaro, Di~Carlo, Dolci, Elias-Rosa, Frandsen, Fynbo, Hopp, Larionov, Laursen, Mazzali, Navasardyan, Ries, Riffeser, Rizzi, Tsvetkov, Turatto, \& Wilke}]{hunter_extensive_2009}
Hunter, D.~J., Valenti, S., Kotak, R., {et~al.} 2009, \bibinfo{title}{Extensive optical and near-infrared observations of the nearby, narrow-lined type {Ic} {SN} 2007gr: days 5 to 415,} Astronomy and Astrophysics, 508, 371, \dodoi{10.1051/0004-6361/200912896}

\bibitem[{J.~D. Hunter(2007)Hunter}]{Matplotlib_Hunter2007}
Hunter, J.~D. 2007, \bibinfo{title}{Matplotlib: A 2D Graphics Environment,} Computing in Science and Engineering, 9, 90, \dodoi{10.1109/MCSE.2007.55}

\bibitem[{P. Höflich {et~al.}(2002)Höflich, Gerardy, Fesen, \& Sakai}]{hoflich_infrared_2002}
Höflich, P., Gerardy, C.~L., Fesen, R.~A., \& Sakai, S. 2002, \bibinfo{title}{Infrared {Spectra} of the {Subluminous} {Type} {Ia} {Supernova} {SN} 1999by,} The Astrophysical Journal, 568, 791, \dodoi{10.1086/339063}

\bibitem[{K. Iwamoto {et~al.}(1994)Iwamoto, Nomoto, Höflich, Yamaoka, Kumagai, \& Shigeyama}]{iwamoto_theoretical_1994}
Iwamoto, K., Nomoto, K., Höflich, P., {et~al.} 1994, \bibinfo{title}{Theoretical {Light} {Curves} for the {Type} {IC} {Supernova} {SN} {1994I},} The Astrophysical Journal, 437, L115, \dodoi{10.1086/187696}

\bibitem[{S. Karthik~Yadavalli {et~al.}(2025)Karthik~Yadavalli, Villar, Polin, Woosley, Drout, \& Pikus}]{karthik_yadavalli_radiative_2025}
Karthik~Yadavalli, S., Villar, V.~A., Polin, A., {et~al.} 2025, \bibinfo{title}{Radiative {Transfer} {Modeling} of {Stripped}-{Envelope} {Supernovae}. {I}: {A} {Grid} for {Ejecta} {Parameter} {Inference},} arXiv, \dodoi{10.48550/arXiv.2507.10648}

\bibitem[{W. Kerzendorf {et~al.}(2022)Kerzendorf, Chen, O'Brien, Buchner, \& van~der Smagt}]{kerzendorf_probabilistic_2022}
Kerzendorf, W., Chen, N., O'Brien, J., Buchner, J., \& van~der Smagt, P. 2022, \bibinfo{title}{Probabilistic {Dalek} -- {Emulator} framework with probabilistic prediction for supernova tomography,}, Tech. rep.
\newblock \url{https://ui.adsabs.harvard.edu/abs/2022arXiv220909453K}

\bibitem[{W. Kerzendorf {et~al.}(2025)Kerzendorf, Sim, Vogl, Williamson, Pássaro, Flörs, Camacho, Jančauskas, Harpole, Nöbauer, Lietzau, Mishin, Tsamis, Boyle, Shingles, Gupta, Desai, Klauser, Beaujean, Suban-Loewen, Heringer, Barna, Gautam, Fullard, Arya, Smith, Cawley, Singhal, Shields, O'Brien, Barbosa, Sondhi, Yu, Patel, Shields, Varanasi, Rathi, Chitchyan, Gillanders, Singh, Savel, Gupta, Reinecke, Holas, Eweis, Bylund, Black, Bentil, Kumar, Eguren, Kumar, Bartnik, Alam, Magee, Dutta, Srivastava, Varma~Buddaraju, Visser, Daksh, Lu, Livneh, Kambham, Roldan, Bhakar, Mishra, Rajagopalan, Reichenbach, Jain, Actions, Floers, Gupta, Chaumal, Brar, Singh, Kowalski, Patidar, Matsumura, Selsing, Sofiatti, Talegaonkar, Kumar, Sharma, Buchner, Yap, Martinez, Truong, Zingale, Sandler, Zaheer, Sarafina, Dasgupta, Patra, Singh~Rathore, Patel, Volodin, Venkat, Prasad, Gupta, Lemoine, Wahi, Aggarwal, Chen, Kolliboyina, PATIDAR, Nayak~U, Kumar, \& Kharkar}]{kerzendorf_tardis-sntardis_2025}
Kerzendorf, W., Sim, S., Vogl, C., {et~al.} 2025, \bibinfo{title}{tardis-sn/tardis: {TARDIS} v2025.03.23,} Zenodo, \dodoi{10.5281/zenodo.15069852}

\bibitem[{W.~E. Kerzendorf \& S.~A. Sim(2014)Kerzendorf \& Sim}]{kerzendorf_spectral_2014}
Kerzendorf, W.~E., \& Sim, S.~A. 2014, \bibinfo{title}{A spectral synthesis code for rapid modelling of supernovae,} Monthly Notices of the Royal Astronomical Society, 440, 387, \dodoi{10.1093/mnras/stu055}

\bibitem[{W.~E. Kerzendorf {et~al.}(2021)Kerzendorf, Vogl, Buchner, Contardo, Williamson, \& van~der Smagt}]{kerzendorf_dalek_2021}
Kerzendorf, W.~E., Vogl, C., Buchner, J., {et~al.} 2021, \bibinfo{title}{Dalek: {A} {Deep} {Learning} {Emulator} for {TARDIS},} The Astrophysical Journal, 910, L23, \dodoi{10.3847/2041-8213/abeb1b}

\bibitem[{D.~P. Kingma \& J. Ba(2017)Kingma \& Ba}]{kingma_adam_2017}
Kingma, D.~P., \& Ba, J. 2017, \bibinfo{title}{Adam: {A} {Method} for {Stochastic} {Optimization},} arXiv, \dodoi{10.48550/arXiv.1412.6980}

\bibitem[{C. Kozma \& C. Fransson(1992)Kozma \& Fransson}]{kozma_gamma-ray_1992}
Kozma, C., \& Fransson, C. 1992, \bibinfo{title}{Gamma-{Ray} {Deposition} and {Nonthermal} {Excitation} in {Supernovae},} The Astrophysical Journal, 390, 602, \dodoi{10.1086/171311}

\bibitem[{H. Kumar {et~al.}(2025{\natexlab{a}})Kumar, Berger, Blanchard, Hiramatsu, Gomez, Gagliano, Andrews, Bostroem, Farah, Howell, \& McCully}]{kumar_detection_2025}
Kumar, H., Berger, E., Blanchard, P.~K., {et~al.} 2025{\natexlab{a}}, \bibinfo{title}{A {Detection} of {Helium} in the {Bright} {Superluminous} {Supernova} {SN} 2024rmj,} The Astrophysical Journal, 992, 122, \dodoi{10.3847/1538-4357/adff7e}

\bibitem[{H. Kumar {et~al.}(2025{\natexlab{b}})Kumar, Berger, Blanchard, Gomez, Hiramatsu, Andrews, Bostroem, Dong, Farah, Padilla~Gonzalez, Howell, McCully, Mehta, Newsome, Ravi, \& Terreran}]{kumar_near-infrared_2025}
Kumar, H., Berger, E., Blanchard, P.~K., {et~al.} 2025{\natexlab{b}}, \bibinfo{title}{A {Near}-infrared {Search} for {Helium} in the {Superluminous} {Supernova} {SN} 2024ahr,} The Astrophysical Journal, 987, 127, \dodoi{10.3847/1538-4357/addc67}

\bibitem[{S. {Kumar} {et~al.}(2026){Kumar}, {Baer-Way}, {Ravi}, {Modjaz}, {Chandra}, {Valenti}, {Kwok}, {Tinyanont}, {Foley}, {Howell}, {Hiramatsu}, {Andrews}, {Bostroem}, {Christy}, {Franz}, {Hsu}, {Pearson}, {Sand}, {Shrestha}, {Smith}, \& {Subrayan}}]{Kumar26_CaStrong}
{Kumar}, S., {Baer-Way}, R., {Ravi}, A.~P., {et~al.} 2026, \bibinfo{title}{{A multiwavelength view of the nearby Calcium-Strong Transient SN 2025coe in the X-Ray, Near-Infrared, and Radio Wavebands},} arXiv e-prints, arXiv:2601.19018, \dodoi{10.48550/arXiv.2601.19018}

\bibitem[{E. Laplace {et~al.}(2021)Laplace, Justham, Renzo, Götberg, Farmer, Vartanyan, \& de~Mink}]{laplace_different_2021}
Laplace, E., Justham, S., Renzo, M., {et~al.} 2021, \bibinfo{title}{Different to the core: {The} pre-supernova structures of massive single and binary-stripped stars,} Astronomy and Astrophysics, 656, A58, \dodoi{10.1051/0004-6361/202140506}

\bibitem[{Y.-Q. Liu {et~al.}(2016)Liu, Modjaz, Bianco, \& Graur}]{liu_analyzing_2016}
Liu, Y.-Q., Modjaz, M., Bianco, F.~B., \& Graur, O. 2016, \bibinfo{title}{Analyzing the {Largest} {Spectroscopic} {Data} {Set} of {Stripped} {Supernovae} to {Improve} {Their} {Identifications} and {Constrain} {Their} {Progenitors},} The Astrophysical Journal, 827, 90, \dodoi{10.3847/0004-637X/827/2/90}

\bibitem[{J. Lu {et~al.}(2025)Lu, Barker, Goldberg, Kerzendorf, Modjaz, Couch, Shields, \& Fullard}]{lu_physics-driven_2025}
Lu, J., Barker, B.~L., Goldberg, J., {et~al.} 2025, \bibinfo{title}{Physics-driven {Explosions} of {Stripped} {High}-mass {Stars}: {Synthetic} {Light} {Curves} and {Spectra} of {Stripped}-envelope {Supernovae} with {Broad} {Light} {Curves},} The Astrophysical Journal, 979, 148, \dodoi{10.3847/1538-4357/ada26d}

\bibitem[{L.~B. Lucy(1991)Lucy}]{lucy_nonthermal_1991}
Lucy, L.~B. 1991, \bibinfo{title}{Nonthermal {Excitation} of {Helium} in {Type} {Ib} {Supernovae},} The Astrophysical Journal, 383, 308, \dodoi{10.1086/170787}

\bibitem[{J.-Z. Ma {et~al.}(2025)Ma, Farmer, de~Mink, \& Laplace}]{ma_carbon_2025}
Ma, J.-Z., Farmer, R., de~Mink, S.~E., \& Laplace, E. 2025, \bibinfo{title}{Carbon from massive binary-stripped stars over cosmic time: effect of metallicity,} arXiv, \dodoi{10.48550/arXiv.2505.02918}

\bibitem[{P.~A. Mazzali {et~al.}(2017)Mazzali, Sauer, Pian, Deng, Prentice, Ben~Ami, Taubenberger, \& Nomoto}]{mazzali_modelling_2017}
Mazzali, P.~A., Sauer, D.~N., Pian, E., {et~al.} 2017, \bibinfo{title}{Modelling the {Type} {Ic} {SN} 2004aw: a moderately energetic explosion of a massive {C}+{O} star without a {GRB},} Monthly Notices of the Royal Astronomical Society, 469, 2498, \dodoi{10.1093/mnras/stx992}

\bibitem[{ {M}c{K}inney {W}es(2010){M}c{K}inney {W}es}]{pandas_paper}
{M}c{K}inney {W}es. 2010, in {P}roceedings of the 9th {P}ython in {S}cience {C}onference, ed. {S}t\'efan van~der {W}alt \& {J}arrod {M}illman, 56 -- 61, \dodoi{10.25080/Majora-92bf1922-00a}

\bibitem[{J. Millard {et~al.}(1999)Millard, Branch, Baron, Hatano, Fisher, Filippenko, Kirshner, Challis, Fransson, Panagia, Phillips, Sonneborn, Suntzeff, Wagoner, \& Wheeler}]{millard_direct_1999}
Millard, J., Branch, D., Baron, E., {et~al.} 1999, \bibinfo{title}{Direct {Analysis} of {Spectra} of the {Type} {IC} {Supernova} {SN} {1994I},} The Astrophysical Journal, 527, 746, \dodoi{10.1086/308108}

\bibitem[{M. Modjaz {et~al.}(2019)Modjaz, Gutiérrez, \& Arcavi}]{modjaz_new_2019}
Modjaz, M., Gutiérrez, C.~P., \& Arcavi, I. 2019, \bibinfo{title}{New regimes in the observation of core-collapse supernovae,} Nature Astronomy, 3, 717, \dodoi{10.1038/s41550-019-0856-2}

\bibitem[{M. Modjaz {et~al.}(2016)Modjaz, Liu, Bianco, \& Graur}]{modjaz_spectral_2016}
Modjaz, M., Liu, Y.~Q., Bianco, F.~B., \& Graur, O. 2016, \bibinfo{title}{The {Spectral} {SN}-{GRB} {Connection}: {Systematic} {Spectral} {Comparisons} between {Type} {Ic} {Supernovae} and {Broad}-lined {Type} {Ic} {Supernovae} with and without {Gamma}-{Ray} {Bursts},} The Astrophysical Journal, 832, 108, \dodoi{10.3847/0004-637X/832/2/108}

\bibitem[{M. Modjaz {et~al.}(2009)Modjaz, Li, Butler, Chornock, Perley, Blondin, Bloom, Filippenko, Kirshner, Kocevski, Poznanski, Hicken, Foley, Stringfellow, Berlind, Barrado~y Navascues, Blake, Bouy, Brown, Challis, Chen, de~Vries, Dufour, Falco, Friedman, Ganeshalingam, Garnavich, Holden, Illingworth, Lee, Liebert, Marion, Olivier, Prochaska, Silverman, Smith, Starr, Steele, Stockton, Williams, \& Wood-Vasey}]{modjaz_shock_2009}
Modjaz, M., Li, W., Butler, N., {et~al.} 2009, \bibinfo{title}{From {Shock} {Breakout} to {Peak} and {Beyond}: {Extensive} {Panchromatic} {Observations} of the {Type} {Ib} {Supernova} {2008D} {Associated} with {Swift} {X}-ray {Transient} 080109,} The Astrophysical Journal, 702, 226, \dodoi{10.1088/0004-637X/702/1/226}

\bibitem[{J.~T. O'Brien {et~al.}(2021)O'Brien, Kerzendorf, Fullard, Williamson, Pakmor, Buchner, Hachinger, Vogl, Gillanders, Flörs, \& van~der Smagt}]{obrien_probabilistic_2021}
O'Brien, J.~T., Kerzendorf, W.~E., Fullard, A., {et~al.} 2021, \bibinfo{title}{Probabilistic {Reconstruction} of {Type} {Ia} {Supernova} {SN} 2002bo,} The Astrophysical Journal, 916, L14, \dodoi{10.3847/2041-8213/ac1173}

\bibitem[{J.~T. O'Brien {et~al.}(2023)O'Brien, Kerzendorf, Fullard, Pakmor, Buchner, Vogl, Chen, van~der Smagt, Williamson, \& Singhal}]{obrien_1991t-like_2023}
O'Brien, J.~T., Kerzendorf, W.~E., Fullard, A., {et~al.} 2023, \bibinfo{title}{{1991T}-{Like} {Type} {Ia} {Supernovae} as an {Extension} of the {Normal} {Population},} \dodoi{10.48550/arXiv.2306.08137}

\bibitem[{J.~T. O'Brien {et~al.}(2024)O'Brien, Kerzendorf, Fullard, Pakmor, Buchner, Vogl, Chen, van~der Smagt, Williamson, \& Singhal}]{obrien_1991t-like_2024}
O'Brien, J.~T., Kerzendorf, W.~E., Fullard, A., {et~al.} 2024, \bibinfo{title}{{1991T}-{Like} {Type} {Ia} {Supernovae} as an {Extension} of the {Normal} {Population},} The Astrophysical Journal, 964, 137, \dodoi{10.3847/1538-4357/ad2358}

\bibitem[{T. pandas~development team(2020)pandas~development team}]{pandas_software}
pandas~development team, T. 2020, \bibinfo{title}{pandas-dev/pandas: Pandas,}, latest Zenodo, \dodoi{10.5281/zenodo.3509134}

\bibitem[{A. Paszke {et~al.}(2019)Paszke, Gross, Massa, Lerer, Bradbury, Chanan, Killeen, Lin, Gimelshein, Antiga, Desmaison, Köpf, Yang, DeVito, Raison, Tejani, Chilamkurthy, Steiner, Fang, Bai, \& Chintala}]{paszke_pytorch_2019}
Paszke, A., Gross, S., Massa, F., {et~al.} 2019, \bibinfo{title}{{PyTorch}: {An} {Imperative} {Style}, {High}-{Performance} {Deep} {Learning} {Library},} arXiv, \dodoi{10.48550/arXiv.1912.01703}

\bibitem[{F. Patat {et~al.}(2001)Patat, Cappellaro, Danziger, Mazzali, Sollerman, Augusteijn, Brewer, Doublier, Gonzalez, Hainaut, Lidman, Leibundgut, Nomoto, Nakamura, Spyromilio, Rizzi, Turatto, Walsh, Galama, van Paradijs, Kouveliotou, Vreeswijk, Frontera, Masetti, Palazzi, \& Pian}]{patat_metamorphosis_2001}
Patat, F., Cappellaro, E., Danziger, J., {et~al.} 2001, \bibinfo{title}{The {Metamorphosis} of {SN} 1998bw,} The Astrophysical Journal, 555, 900, \dodoi{10.1086/321526}

\bibitem[{F. Pedregosa {et~al.}(2011)Pedregosa, Varoquaux, Gramfort, Michel, Thirion, Grisel, Blondel, Prettenhofer, Weiss, Dubourg, Vanderplas, Passos, Cournapeau, Brucher, Perrot, \& Duchesnay}]{scikit-learn}
Pedregosa, F., Varoquaux, G., Gramfort, A., {et~al.} 2011, \bibinfo{title}{Scikit-learn: Machine Learning in {P}ython,} Journal of Machine Learning Research, 12, 2825

\bibitem[{M.~M. {Phillips} {et~al.}(2019){Phillips}, {Contreras}, {Hsiao}, {Morrell}, {Burns}, {Stritzinger}, {Ashall}, {Freedman}, {Hoeflich}, {Persson}, {Piro}, {Suntzeff}, {Uddin}, {Anais}, {Baron}, {Busta}, {Campillay}, {Castell{\'o}n}, {Corco}, {Diamond}, {Gall}, {Gonzalez}, {Holmbo}, {Krisciunas}, {Roth}, {Ser{\'o}n}, {Taddia}, {Torres}, {Anderson}, {Baltay}, {Folatelli}, {Galbany}, {Goobar}, {Hadjiyska}, {Hamuy}, {Kasliwal}, {Lidman}, {Nugent}, {Perlmutter}, {Rabinowitz}, {Ryder}, {Schmidt}, {Shappee}, \& {Walker}}]{Phillips2019}
{Phillips}, M.~M., {Contreras}, C., {Hsiao}, E.~Y., {et~al.} 2019, \bibinfo{title}{{Carnegie Supernova Project-II: Extending the Near-infrared Hubble Diagram for Type Ia Supernovae to z{\nbsp}{\sim}{\nbsp}0.1},} \pasp, 131, 014001, \dodoi{10.1088/1538-3873/aae8bd}

\bibitem[{A.~L. Piro \& V.~S. Morozova(2014)Piro \& Morozova}]{piro_transparent_2014}
Piro, A.~L., \& Morozova, V.~S. 2014, \bibinfo{title}{Transparent {Helium} in {Stripped} {Envelope} {Supernovae},} The Astrophysical Journal, 792, L11, \dodoi{10.1088/2041-8205/792/1/L11}

\bibitem[{P. Podsiadlowski {et~al.}(1992)Podsiadlowski, Joss, \& Hsu}]{podsiadlowski_presupernova_1992}
Podsiadlowski, P., Joss, P.~C., \& Hsu, J. J.~L. 1992, \bibinfo{title}{Presupernova {Evolution} in {Massive} {Interacting} {Binaries},} The Astrophysical Journal, 391, 246, \dodoi{10.1086/171341}

\bibitem[{H. Sana {et~al.}(2012)Sana, de~Mink, de~Koter, Langer, Evans, Gieles, Gosset, Izzard, Le~Bouquin, \& Schneider}]{sana_binary_2012}
Sana, H., de~Mink, S.~E., de~Koter, A., {et~al.} 2012, \bibinfo{title}{Binary {Interaction} {Dominates} the {Evolution} of {Massive} {Stars},} Science, 337, 444, \dodoi{10.1126/science.1223344}

\bibitem[{D.~N. Sauer {et~al.}(2006)Sauer, Mazzali, Deng, Valenti, Nomoto, \& Filippenko}]{sauer_properties_2006}
Sauer, D.~N., Mazzali, P.~A., Deng, J., {et~al.} 2006, \bibinfo{title}{The properties of the `standard' {Type} {Ic} supernova {1994I} from spectral models,} Monthly Notices of the Royal Astronomical Society, 369, 1939, \dodoi{10.1111/j.1365-2966.2006.10438.x}

\bibitem[{M. Shahbandeh {et~al.}(2022)Shahbandeh, Hsiao, Ashall, Teffs, Hoeflich, Morrell, Phillips, Anderson, Baron, Burns, Contreras, Davis, Diamond, Folatelli, Galbany, Gall, Hachinger, Holmbo, Karamehmetoglu, Kasliwal, Kirshner, Krisciunas, Kumar, Lu, Marion, Mazzali, Piro, Sand, Stritzinger, Suntzeff, Taddia, \& Uddin}]{shahbandeh_carnegie_2022}
Shahbandeh, M., Hsiao, E.~Y., Ashall, C., {et~al.} 2022, \bibinfo{title}{Carnegie {Supernova} {Project}-{II}: {Near}-infrared {Spectroscopy} of {Stripped}-envelope {Core}-collapse {Supernovae},} The Astrophysical Journal, 925, 175, \dodoi{10.3847/1538-4357/ac4030}

\bibitem[{R.~A. Simcoe {et~al.}(2013)Simcoe, Burgasser, Schechter, Fishner, Bernstein, Bigelow, Pipher, Forrest, McMurtry, Smith, \& Bochanski}]{simcoe_fire_2013}
Simcoe, R.~A., Burgasser, A.~J., Schechter, P.~L., {et~al.} 2013, \bibinfo{title}{{FIRE}: {A} {Facility} {Class} {Near}-{Infrared} {Echelle} {Spectrometer} for the {Magellan} {Telescopes},} Publications of the Astronomical Society of the Pacific, 125, 270, \dodoi{10.1086/670241}

\bibitem[{J. Skilling(2004)Skilling}]{skilling_nested_2004}
Skilling, J. 2004, in , 395--405, \dodoi{10.1063/1.1835238}

\bibitem[{M. Solar {et~al.}(2024)Solar, Michałowski, Nadolny, Galbany, Hjorth, Zapartas, Sollerman, Hunt, Klose, Koprowski, Leśniewska, Małkowski, Nicuesa~Guelbenzu, Ryzhov, Savaglio, Schady, Schulze, de~Ugarte~Postigo, Vergani, Watson, \& Wróblewski}]{solar_binary_2024}
Solar, M., Michałowski, M.~J., Nadolny, J., {et~al.} 2024, \bibinfo{title}{Binary progenitor systems for {Type} {Ic} supernovae,} Nature Communications, 15, 7667, \dodoi{10.1038/s41467-024-51863-z}

\bibitem[{N.-C. Sun {et~al.}(2023)Sun, Maund, \& Crowther}]{sun_uv_2023}
Sun, N.-C., Maund, J.~R., \& Crowther, P.~A. 2023, \bibinfo{title}{A {UV} census of the environments of stripped-envelope supernovae,} Monthly Notices of the Royal Astronomical Society, 521, 2860, \dodoi{10.1093/mnras/stad690}

\bibitem[{S. Taubenberger {et~al.}(2006)Taubenberger, Pastorello, Mazzali, Valenti, Pignata, Sauer, Arbey, Bärnbantner, Benetti, Della~Valle, Deng, Elias-Rosa, Filippenko, Foley, Goobar, Kotak, Li, Meikle, Mendez, Patat, Pian, Ries, Ruiz-Lapuente, Salvo, Stanishev, Turatto, \& Hillebrandt}]{taubenberger_sn_2006}
Taubenberger, S., Pastorello, A., Mazzali, P.~A., {et~al.} 2006, \bibinfo{title}{{SN} 2004aw: confirming diversity of {Type} {Ic} supernovae,} Monthly Notices of the Royal Astronomical Society, 371, 1459, \dodoi{10.1111/j.1365-2966.2006.10776.x}

\bibitem[{J. Teffs {et~al.}(2020{\natexlab{a}})Teffs, Ertl, Mazzali, Hachinger, \& Janka}]{teffs_how_2020}
Teffs, J., Ertl, T., Mazzali, P., Hachinger, S., \& Janka, H.~T. 2020{\natexlab{a}}, \bibinfo{title}{How much {H} and {He} is 'hidden' in {SNe} {Ib}/c? - {II}. {Intermediate}-mass objects: a 22-={M}⊙ progenitor case study,} Monthly Notices of the Royal Astronomical Society, 499, 730, \dodoi{10.1093/mnras/staa2549}

\bibitem[{J. Teffs {et~al.}(2020{\natexlab{b}})Teffs, Ertl, Mazzali, Hachinger, \& Janka}]{teffs_type_2020}
Teffs, J., Ertl, T., Mazzali, P., Hachinger, S., \& Janka, T. 2020{\natexlab{b}}, \bibinfo{title}{Type {Ic} supernova of a 22 {M}⊙ progenitor,} Monthly Notices of the Royal Astronomical Society, 492, 4369, \dodoi{10.1093/mnras/staa123}

\bibitem[{J.~J. Teffs {et~al.}(2021)Teffs, Prentice, Mazzali, \& Ashall}]{teffs_observations_2021}
Teffs, J.~J., Prentice, S.~J., Mazzali, P.~A., \& Ashall, C. 2021, \bibinfo{title}{Observations and spectral modelling of the narrow-lined {Type} {Ic} {SN} 2017ein,} Monthly Notices of the Royal Astronomical Society, 502, 3829, \dodoi{10.1093/mnras/stab258}

\bibitem[{B.~T.-H. Tsang {et~al.}(2020)Tsang, Goldberg, Bildsten, \& Kasen}]{tsang_comparing_2020}
Tsang, B. T.-H., Goldberg, J.~A., Bildsten, L., \& Kasen, D. 2020, \bibinfo{title}{Comparing {Moment}-based and {Monte} {Carlo} {Methods} of {Radiation} {Transport} {Modeling} for {Type} {II}-{Plateau} {Supernova} {Light} {Curves},} The Astrophysical Journal, 898, 29, \dodoi{10.3847/1538-4357/ab989d}

\bibitem[{S. Valenti {et~al.}(2008)Valenti, Elias-Rosa, Taubenberger, Stanishev, Agnoletto, Sauer, Cappellaro, Pastorello, Benetti, Riffeser, Hopp, Navasardyan, Tsvetkov, Lorenzi, Patat, Turatto, Barbon, Ciroi, Di~Mille, Frandsen, Fynbo, Laursen, \& Mazzali}]{valenti_carbon-rich_2008}
Valenti, S., Elias-Rosa, N., Taubenberger, S., {et~al.} 2008, \bibinfo{title}{The {Carbon}-rich {Type} {Ic} {SN} 2007gr: {The} {Photospheric} {Phase},} The Astrophysical Journal, 673, L155, \dodoi{10.1086/527672}

\bibitem[{P. Virtanen {et~al.}(2020)Virtanen, Gommers, Oliphant, Haberland, Reddy, Cournapeau, Burovski, Peterson, Weckesser, Bright, {van der Walt}, Brett, Wilson, Millman, Mayorov, Nelson, Jones, Kern, Larson, Carey, Polat, Feng, Moore, {VanderPlas}, Laxalde, Perktold, Cimrman, Henriksen, Quintero, Harris, Archibald, Ribeiro, Pedregosa, {van Mulbregt}, \& {SciPy 1.0 Contributors}}]{2020SciPy-NMeth}
Virtanen, P., Gommers, R., Oliphant, T.~E., {et~al.} 2020, \bibinfo{title}{{{SciPy} 1.0: Fundamental Algorithms for Scientific Computing in Python},} Nature Methods, 17, 261, \dodoi{10.1038/s41592-019-0686-2}

\bibitem[{C. Vogl {et~al.}(2020)Vogl, Kerzendorf, Sim, Noebauer, Lietzau, \& Hillebrandt}]{vogl_spectral_2020}
Vogl, C., Kerzendorf, W.~E., Sim, S.~A., {et~al.} 2020, \bibinfo{title}{Spectral modeling of type {II} supernovae. {II}. {A} machine-learning approach to quantitative spectroscopic analysis,} Astronomy and Astrophysics, 633, A88, \dodoi{10.1051/0004-6361/201936137}

\bibitem[{C. Vogl {et~al.}(2024)Vogl, Taubenberger, Csörnyei, Leibundgut, Kerzendorf, Sim, Blondin, Flörs, Holas, Shields, Spyromilio, Suyu, \& Hillebrandt}]{vogl_no_2024}
Vogl, C., Taubenberger, S., Csörnyei, G., {et~al.} 2024, \bibinfo{title}{No rungs attached: {A} distance-ladder free determination of the {Hubble} constant through type {II} supernova spectral modelling,} \dodoi{10.48550/arXiv.2411.04968}

\bibitem[{J.~C. Wheeler {et~al.}(1994)Wheeler, Harkness, Clocchiatti, Benetti, Brotherton, Depoy, \& Elias}]{wheeler_sn_1994}
Wheeler, J.~C., Harkness, R.~P., Clocchiatti, A., {et~al.} 1994, \bibinfo{title}{{SN} {1994I} in {M51} and the nature of type {IBC} supernovae.,} The Astrophysical Journal, 436, L135, \dodoi{10.1086/187651}

\bibitem[{J.~C. Wheeler {et~al.}(1998)Wheeler, Höflich, Harkness, \& Spyromilio}]{wheeler_explosion_1998}
Wheeler, J.~C., Höflich, P., Harkness, R.~P., \& Spyromilio, J. 1998, \bibinfo{title}{Explosion {Diagnostics} of {Type} {IA} {Supernovae} from {Early} {Infrared} {Spectra},} The Astrophysical Journal, 496, 908, \dodoi{10.1086/305427}

\bibitem[{J.~C. Wheeler \& R. Levreault(1985)Wheeler \& Levreault}]{wheeler_peculiar_1985}
Wheeler, J.~C., \& Levreault, R. 1985, \bibinfo{title}{The peculiar type {I} supernova in {NGC} 991.,} The Astrophysical Journal, 294, L17, \dodoi{10.1086/184500}

\bibitem[{M. Williamson {et~al.}(2021)Williamson, Kerzendorf, \& Modjaz}]{williamson_modeling_2021}
Williamson, M., Kerzendorf, W., \& Modjaz, M. 2021, \bibinfo{title}{Modeling {Type} {Ic} {Supernovae} with {TARDIS}: {Hidden} {Helium} in {SN} {1994I}?} The Astrophysical Journal, 908, 150, \dodoi{10.3847/1538-4357/abd244}

\bibitem[{M. Williamson {et~al.}(2023)Williamson, Vogl, Modjaz, Kerzendorf, Singhal, Boland, Burke, Chen, Hiramatsu, Galbany, Gonzalez, Howell, Jha, Kwok, McCully, Newsome, Pellegrino, Rho, Terreran, \& Wang}]{williamson_sn_2023}
Williamson, M., Vogl, C., Modjaz, M., {et~al.} 2023, \bibinfo{title}{{SN} 2019ewu: {A} {Peculiar} {Supernova} with {Early} {Strong} {Carbon} and {Weak} {Oxygen} {Features} from a {New} {Sample} of {Young} {SN} {Ic} {Spectra},} The Astrophysical Journal, 944, L49, \dodoi{10.3847/2041-8213/acb549}

\bibitem[{S.~E. Woosley {et~al.}(1993)Woosley, Langer, \& Weaver}]{woosley_evolution_1993}
Woosley, S.~E., Langer, N., \& Weaver, T.~A. 1993, \bibinfo{title}{The {Evolution} of {Massive} {Stars} {Including} {Mass} {Loss}: {Presupernova} {Models} and {Explosion},} The Astrophysical Journal, 411, 823, \dodoi{10.1086/172886}

\bibitem[{S.-C. Yoon(2015)Yoon}]{yoon_evolutionary_2015}
Yoon, S.-C. 2015, \bibinfo{title}{Evolutionary {Models} for {Type} {Ib}/c {Supernova} {Progenitors},} Publications of the Astronomical Society of Australia, 32, e015, \dodoi{10.1017/pasa.2015.16}

\bibitem[{S.~C. Yoon {et~al.}(2010)Yoon, Woosley, \& Langer}]{yoon_type_2010}
Yoon, S.~C., Woosley, S.~E., \& Langer, N. 2010, \bibinfo{title}{Type {Ib}/c {Supernovae} in {Binary} {Systems}. {I}. {Evolution} and {Properties} of the {Progenitor} {Stars},} The Astrophysical Journal, 725, 940, \dodoi{10.1088/0004-637X/725/1/940}

\bibitem[{E. Zapartas {et~al.}(2025)Zapartas, Fox, Su, Souropanis, Drout, Rocha, van Dyk, Williams, Briel, Renzo, Andrews, Fragos, Gossage, Kruckow, Liotine, Ryder, Srivastava, \& Teng}]{zapartas_demographics_2025}
Zapartas, E., Fox, O.~D., Su, J., {et~al.} 2025, \bibinfo{title}{The {Demographics} of {Binary} {Companions} to {Stripped}-{Envelope} {Supernovae}: {Confronting} {Observations} with {Population} {Synthesis},} arXiv, \dodoi{10.48550/arXiv.2508.12677}

\bibitem[{J. Zhang {et~al.}(2018)Zhang, Wang, Vinkó, Wheeler, Chang, Yang, Wang, Zhai, Rui, Mo, Zhang, Zhang, Wang, Mao, Wang, Yi, Xin, Li, Lun, Lu, Sai, Zheng, Zhang, Zhou, \& Bai}]{zhang_optical_2018}
Zhang, J., Wang, X., Vinkó, J., {et~al.} 2018, \bibinfo{title}{Optical {Observations} of the {Young} {Type} {Ic} {Supernova} {SN} {2014L} in {M99},} The Astrophysical Journal, 863, 109, \dodoi{10.3847/1538-4357/aaceaf}

\end{thebibliography}
\bibliographystyle{aasjournalv7}

%%%%%%%%%%%%%%%%%%%%%%%%%%%%%%%%%%%%%%%%%%%%%%%%%%%%%%%%%%%%%%%%%%%%%%%%%%%%%%%%
\appendix
\restartappendixnumbering 

\section{Emulator Architecture}\label{appendix: emulator achitechture}
%%%%%% Architecture design
We adopt the network design of \texttt{Probabilistic Dalek} \citep{kerzendorf_probabilistic_2022}, which was tuned on synthetic \tardis spectra and consists of five hidden layers between the input and output layers, each containing 400 neurons and followed by a \texttt{Softplus} activation function \citep{dugas2000incorporating}.
Local connections are implemented by concatenating the output of each layer with the input to the subsequent layer, which mitigates vanishing gradients and reduces feature redundancy \citep{he_deep_2016, huang_densely_2017}.
The output layer is structured to return both the mean prediction of the normalized spectral features obtained through a \texttt{linear} activation, and the associated predictive uncertainty modeled by applying a \texttt{Softplus} activation to the corresponding output node.
In this work, training is performed using mini-batch gradient descent with a batch size of 1024 and a total epoch of \num{50000}, optimized with the \texttt{Adam} algorithm \citep{kingma_adam_2017} and a learning rate of $3e^{-4}$. 
We select the final model based on validation dataset performance, adopting the checkpoint that achieves the lowest validation loss.

%%%%%%%%%%%%%%%%%%%%%%%%%%%%%%%%%%%%%%%%%%%%%%%%%%
\section{Emulator Performance} \label{appendix: emulator performance}
%%%% definition of the meanFE metric
We evaluate the performance of our emulator by quantifying the mean fractional error (meanFE) in flux space, which is the same metric used in previous \tardis-based emulator works \citep{vogl_spectral_2020, kerzendorf_dalek_2021, obrien_probabilistic_2021,kerzendorf_probabilistic_2022,obrien_1991t-like_2024}:
\begin{equation}
\text { meanFE }  =\frac{1}{N} \sum_{i=0}^N \frac{\left|f_i^{\mathrm{emu}}-f_i^{\mathrm{\tardis}}\right|}{f_i^{\mathrm{\tardis}}},
\end{equation}
where N is the dimension of the wavelength grid,  $f_i^{\mathrm{emu}}$ and $f_i^{\mathrm{\tardis}}$ represents the emulator and \tardis luminosity density at the $i$-th wavelength point, respectively. 

\begin{figure}[tbh!]
\centering
\includegraphics[width=\columnwidth]{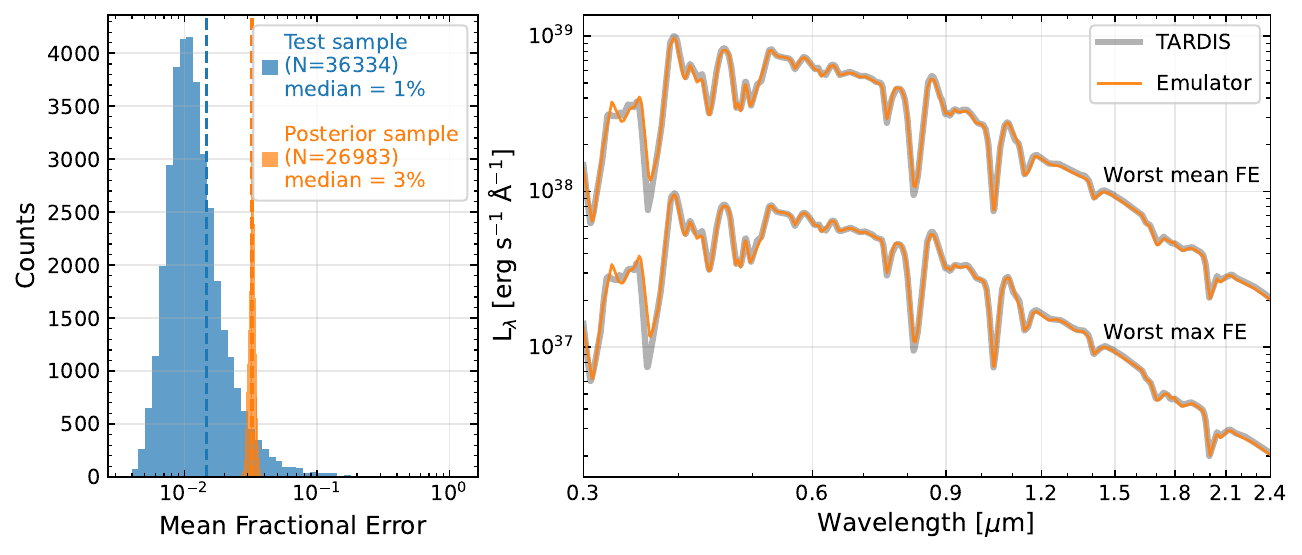}
\caption{\textit{Left}: Distribution of the mean fractional error in flux evaluated using the test dataset and posterior parameter samples. The vertical dashed lines mark the median of each distribution. 
\textit{Right}: The comparison of the \tardis and emulator spectra of two parameter sets that yield the worst fractional errors within the sampled posterior set.}
\label{fig: emulator_performance}
\end{figure}

%%%% report the results of mean FE
Evaluating on the emulator test dataset (\num{36334} samples), the emulator yields a median meanFE of 1\%. 
To quantify the emulator performance within the posterior space of the near-peak spectra of SN~2014L, we randomly choose \num{26983} samples in the resultant posterior distribution and ran through \tardis, obtaining a median meanFE of 3\%.
We present the meanFE distribution evaluated using the emulator test sample and the posterior sample in the left panel of Figure~\ref{fig: emulator_performance}.
In the right panel of Figure~\ref{fig: emulator_performance}, we showcase the comparison between the \tardis and emulator spectra using two parameter sets that yield the worst fractional errors within the sampled posterior set.
The top one demonstrates the parameter set that returns the worst meanFE metric.
The bottom one shows the parameter set that yields the highest maximum fractional error in a given spectrum.

%%%%%%%%%%%%%%%%%%%%%%%%%%%%%%%%%%%%%%%%
%%%%%%%%%%%%%%%%%%%%%%%%%%%%%%%%%%%%%%%%
%%%%%%%%%%%%%%%%%%%%%%%%%%%%%%%%%%%%%%%%
%%%%%%%%%%%%%%%%%%%%%%%%%%%%%%%%%%%%%%%%
%%%%%%%%%%%%%%%%%%%%%%%%%%%%%%%%%%%%%%%%%%%%%%%%%%%%%%%%%%%%%%%%%%%%%%%%%%%%%%%%

\section*{ACKNOWLEDGEMENT}
%%%%%% general 
We thank Marc Williamson for the discussion on the SN~Ic emulator construction. 
We thank Zhujia Zhang for providing the data file of SN~2014L that contains the flux uncertainty.
We thank Jingze Ma for the helpful discussion on the surface C/O ratio.
We thank Johannes Buchner for the guidance on \ultranest.
%%%%%% TARDIS
This research made use of \tardis, a community-developed software package for spectral synthesis in supernovae. 
The development of \tardis received support from GitHub, the Google Summer of Code initiative, and ESA's Summer of Code in Space program. 
\tardis is a fiscally sponsored project of NumFOCUS. 
\tardis makes extensive use of Astropy.
%%%% Computing
This work was supported in part by Michigan State University through computational resources provided by the Institute for Cyber-Enabled Research.
%%%% Grants acknowledgements
J.L. and J.V.S. are supported by grant NSF-2206523.  
W.E.K. acknowledges financial support from NSF-2206523 and NSF-2311323.
M.M. acknowledges support in part from ADAP program grant No. 80NSSC22K0486, from the NSF grant AST-2206657, from the HST GO program HST-GO-16656 and from the National Science Foundation under Cooperative Agreement 2421782 and the Simons Foundation grant MPS-AI-00010515 awarded to the NSF-Simons AI Institute for Cosmic Origins — CosmicAI, https://www.cosmicai.org/.
J.A.G. acknowledges financial support from NASA grant 23-ATP23-0070. The Flatiron Institute is supported by the Simons Foundation.

\software{
\texttt{astropy}\footnote{\url{https://www.astropy.org/}} \citep{astropy:2013,astropy:2018, astropy:2022}, 
\texttt{matplotlib}\footnote{\url{https://matplotlib.org/}} \citep{Matplotlib_Hunter2007},
\texttt{pandas}\footnote{\url{https://pandas.pydata.org/}} \citep{pandas_paper,pandas_software},
\texttt{pytorch}\footnote{\url{https://pytorch.org/}} \citep{paszke_pytorch_2019},
\texttt{pytorch-lighting}\footnote{\url{https://lightning.ai/docs/pytorch/stable/}} \citep{falcon_pytorch_2024},
\texttt{scikit-learn}\footnote{\url{https://scikit-learn.org/stable/}} \citep{scikit-learn},
\texttt{scipy}\footnote{\url{https://scipy.org/}} \citep{2020SciPy-NMeth},
\texttt{tardis}\footnote{\url{https://github.com/tardis-sn/tardis}} (\citealt{kerzendorf_spectral_2014}; version: \citealt{kerzendorf_tardis-sntardis_2025})},
\texttt{ultranest}\footnote{\url{https://github.com/JohannesBuchner/UltraNest}} \citep{buchner_ultranest_2021}.

{\it Contributor Roles:} J.L. led the analysis and writing of the paper. W.E.K. led the project and oversaw the project progress. J.L. and J.T.O led the modeling efforts. J.G. and M.M. contributed to the analysis and interpretation. N.C, E.V., J.V.S., and A.G.F contributed relevant scientific expertise to the project.

\end{CJK*}
\end{document}